\documentclass[journal]{IEEEtran}

\usepackage{graphicx}
\usepackage{multirow}
\usepackage[none]{hyphenat}
\usepackage{float}
\usepackage{subcaption}
\usepackage{dblfloatfix}
\usepackage[cmex10,intlimits]{amsmath}
\usepackage{epsf}
\usepackage{tikz,pgfplots,tikz-3dplot}
\usetikzlibrary{shapes.geometric, arrows}
\usetikzlibrary{decorations.shapes}
\usepackage{bigstrut}
\usepackage[noadjust]{cite}
\usepackage{mathtools}
\usepackage{algorithm}
\usepackage{algpseudocode}
\usepackage{amssymb}
\usepackage{gensymb}
\usepackage{placeins}
\usepackage[font=small,labelfont=bf]{caption}
\graphicspath{{./figs/}}  

\usepackage{multicol}
\usepackage{array}
\newcolumntype{L}[1]{>{\raggedright\let\newline\\\arraybackslash\hspace{0pt}}m{#1}}
\newcolumntype{C}[1]{>{\centering\let\newline\\\arraybackslash\hspace{0pt}}m{#1}}
\newcolumntype{R}[1]{>{\raggedleft\let\newline\\\arraybackslash\hspace{0pt}}m{#1}}

\newcommand{\gv}[1]{\ensuremath{\mbox{\boldmath$ #1 $}}}

\usepackage[T1]{fontenc}
\usepackage{times}
\makeatletter

\def\@IMSauthorblockNAMEstyle{\normalfont\IMSauthorsize}
\def\@IMSauthorblockAFFILstyle{\normalfont\IMSaffilsize}
\def\@IMSauthorblockEMAILstyle{\normalfont\IMSaffilsize}
\def\IMSauthorblockNAME#1{%
\relax\@IMSauthorblockNAMEstyle%
#1%
}%
\def\IMSauthorblockAFFIL#1{%
\relax\@IMSauthorblockAFFILstyle%
\vskip\@IEEEauthorblockAtopspace
#1%
}%
\def\IMSauthorblockEMAIL#1{%
\relax\@IMSauthorblockEMAILstyle%
\vskip\@IEEEauthorblockAtopspace
#1%
}%
\newif\ifIsBlindReviewVersion

\def\@maketitle{\newpage
\bgroup\par\addvspace{0.5\baselineskip}\centering%
\ifCLASSOPTIONtechnote
   {\bfseries\large\@IEEEcompsoconly{\sffamily}\@title\par}\vskip 1.3em{\lineskip .5em\@IEEEcompsoconly{\sffamily}\@author
   \@IEEEspecialpapernotice\par{\@IEEEcompsoconly{\vskip 1.5em\relax
   \@IEEEtitleabstractindextextbox{\@IEEEtitleabstractindextext}\par
   \hfill\@IEEEcompsocdiamondline\hfill\hbox{}\par}}}\relax
\else
   \vskip0.2em{\IMStitlesize\ifCLASSOPTIONtransmag\bfseries\LARGE\fi\@IEEEcompsoconly{\sffamily}\@IEEEcompsocconfonly{\normalfont\normalsize\vskip 2\@IEEEnormalsizeunitybaselineskip
   \bfseries\Large}\@title\par}\vskip1.0em\par
   \ifCLASSOPTIONconference%
      {\@IEEEspecialpapernotice\mbox{}\vskip\@IEEEauthorblockconfadjspace%
       \mbox{}\hfill\begin{@IEEEauthorhalign}\@author\end{@IEEEauthorhalign}\hfill\mbox{}\par}\relax
   \else
      \ifCLASSOPTIONpeerreviewca
         {\@IEEEcompsoconly{\sffamily}\@IEEEspecialpapernotice\mbox{}\vskip\@IEEEauthorblockconfadjspace%
          \mbox{}\hfill\begin{@IEEEauthorhalign}\@author\end{@IEEEauthorhalign}\hfill\mbox{}\par
          {\@IEEEcompsoconly{\vskip 1.5em\relax
           \@IEEEtitleabstractindextextbox{\@IEEEtitleabstractindextext}\par\hfill
           \@IEEEcompsocdiamondline\hfill\hbox{}\par}}}\relax
      \else
         \ifCLASSOPTIONtransmag
           {\@IEEEspecialpapernotice\mbox{}\vskip\@IEEEauthorblockconfadjspace%
            \mbox{}\hfill\begin{@IEEEauthorhalign}\@author\end{@IEEEauthorhalign}\hfill\mbox{}\par
           {\vspace{0.5\baselineskip}\relax\@IEEEtitleabstractindextextbox{\@IEEEtitleabstractindextext}\vspace{-1\baselineskip}\par}}\relax
         \else
           {\lineskip.5em\@IEEEcompsoconly{\sffamily}\sublargesize\@author\@IEEEspecialpapernotice\par
           {\@IEEEcompsoconly{\vskip 1.5em\relax
            \@IEEEtitleabstractindextextbox{\@IEEEtitleabstractindextext}\par\hfill
            \@IEEEcompsocdiamondline\hfill\hbox{}\par}}}\relax
         \fi
      \fi
   \fi
\fi\par\addvspace{0.0\baselineskip}\egroup}

\def\IMStitlesize{\@setfontsize{\IMStitlesize}{18}{21pt}}
\def\IMSauthorsize{\@setfontsize{\IMSauthorsize}{12}{13pt}}
\def\IMSaffilsize{\@setfontsize{\IMSaffilsize}{12}{13pt}}
\def\IMScaptionsize{\@setfontsize{\IMScaptionsize}{8}{9pt}}
\def\IMSbibsize{\@setfontsize{\IMSbibsize}{8}{9pt}}

\def\@IEEEauthorblockNstyle{\IMSauthorsize\@IEEEcompsocnotconfonly{\sffamily}\@IEEEcompsocconfonly{\large}}
\def\@IEEEauthorblockAstyle{\IMSaffilsize\@IEEEcompsocnotconfonly{\sffamily}\@IEEEcompsocconfonly{\itshape}\@IEEEcompsocconfonly{\large}}
\def\@IEEEauthordefaulttextstyle{\IMSauthorsize\@IEEEcompsocnotconfonly{\sffamily}\sublargesize}

\def\thebibliography#1{\section*{\refname}%
    \addcontentsline{toc}{section}{\refname}%
    \IMSbibsize\@IEEEcompsocconfonly{\small}\vskip 0.3\baselineskip plus 0.1\baselineskip minus 0.1\baselineskip
    \list{\@biblabel{\@arabic\c@enumiv}}%
    {\settowidth\labelwidth{\@biblabel{#1}}%
    \leftmargin\labelwidth
    \advance\leftmargin\labelsep\relax
    \itemsep \IEEEbibitemsep\relax
    \usecounter{enumiv}%
    \let\p@enumiv\@empty
    \renewcommand\theenumiv{\@arabic\c@enumiv}}%
    \let\@IEEElatexbibitem\bibitem%
    \def\bibitem{\@IEEEbibitemprefix\@IEEElatexbibitem}%
\def\newblock{\hskip .11em plus .33em minus .07em}%
\ifCLASSOPTIONtechnote\sloppy\clubpenalty4000\widowpenalty4000\interlinepenalty100%
\else\sloppy\clubpenalty4000\widowpenalty4000\interlinepenalty500\fi%
    \sfcode`\.=1000\relax}

\long\def\@makecaption#1#2{%
\ifx\@captype\@IEEEtablestring%
\par\@IEEEtabletopskipstrut
\else
\@IEEEfigurecaptionsepspace
\fi
\setbox\@tempboxa\hbox{\normalfont\IMScaptionsize {#1.}\nobreakspace\nobreakspace #2}%
\ifdim \wd\@tempboxa >\hsize%
\setbox\@tempboxa\hbox{\normalfont\IMScaptionsize {#1.}\nobreakspace\nobreakspace}%
\parbox[t]{\hsize}{\normalfont\IMScaptionsize\noindent\unhbox\@tempboxa#2}%
\else
\ifCLASSOPTIONconference \hbox to\hsize{\normalfont\IMScaptionsize\hfil\box\@tempboxa\hfil}%
\else \hbox to\hsize{\normalfont\IMScaptionsize\box\@tempboxa\hfil}%
\fi\fi
\ifx\@captype\@IEEEtablestring%
\@IEEEtablecaptionsepspace
\else
\fi}

\newlength\tablecaptiontotableskip
\newlength\figuretocaptionskip
\def\@IEEEfigurecaptionsepspace{\vskip\figuretocaptionskip\relax}%
\def\@IEEEtablecaptionsepspace{\vskip\tablecaptiontotableskip\relax}%

\def\abstract{\normalfont%
\@IEEEabskeysecsize\bfseries\textit{\abstractname}\,\bfseries\textit{---}\,%
\@IEEEgobbleleadPARNLSP}%

\def\IEEEkeywords{\normalfont%
\@IEEEabskeysecsize\bfseries\textit{\IEEEkeywordsname}\,\bfseries\textit{---}\,%
\@IEEEgobbleleadPARNLSP}%
\def\endIEEEkeywords{\relax\vspace{0.67ex}%
\par\if@twocolumn\else\endquotation\fi%
\normalsize\normalfont}%

\def\@IEEEauthorblockNtopspace{0ex}
\def\@IEEEauthorblockAtopspace{1mm}
\def\IEEEkeywordsname{Keywords}
\def\subsubsection{\@startsection{subsubsection}{3}{\z@}{1.5ex plus 1.5ex minus 0.5ex}%
{0.7ex plus .5ex minus 0ex}{\normalfont\normalsize\itshape}}%
\def\@seccntformat#1{\csname the#1dis\endcsname\relax}
\def\thesubsectiondis{{\hbox to\parindent{\Alph{subsection}.}}}
\def\thesubsubsectiondis{{\hbox to \parindent{\arabic{subsubsection})}}}
\def\theparagraphdis{{\hbox to \parindent{\alph{paragraph})}}}
\IEEEilabelindentA \parindent
\IEEEilabelindent \IEEEilabelindentA
\IEEEelabelindent \parindent
\IEEEdlabelindent \parindent
\IEEElabelindent \parindent

\newlength\@IMSparindent
\newcommand\IMSdisplayacksection[1]{%
\ifIsBlindReviewVersion%
\noindent\phantom{\parbox[t]{\columnwidth}{\normalbaselines\setlength{\parindent}{\@IMSparindent}{#1}\strut}}
\else%
\noindent\parbox[t]{\columnwidth}{\normalbaselines\setlength{\parindent}{\@IMSparindent}{#1}\strut}%
\fi%
}%

\makeatother

\begin{document}

%
%

        \title{TT-FDTD: Tensor Train Accelerated Three-Dimensional FDTD With Logarithmic Cost of Spatial Operators}
  \author{Chris~Nguyen,~\IEEEmembership{Student Member,~IEEE,}
 and~Vladimir~I.~Okhmatovski,~\IEEEmembership{Senior Member,~IEEE}

  \thanks{Manuscript received July 27, 2026; revised [Month Day, Year].}
  \thanks{C. Nguyen and V. I. Okhmatovski are with the Department of Electrical and Computer Engineering, University of Manitoba, Winnipeg, MB R3T~5V6, Canada (e-mail: nguye73@myumanitoba.ca; vladimir.okhmatovski@umanitoba.ca).}}  
 \maketitle

\begin{abstract}
Quantized tensor-train (QTT) compression is incorporated into a full-vector three-dimensional scattered-field finite-difference time-domain (FDTD) formulation on uniform Yee grids. All six electromagnetic-field components, material-dependent update coefficients, equivalent-current sources, and staggered finite-difference operators are represented in compatible QTT form. Gaussian regularization of voxelized material interfaces is used to reduce the coefficient ranks generated by abrupt dielectric and conductivity transitions. The formulation is evaluated for an anatomically heterogeneous human-head model and a homogeneous dielectric sphere on grids containing up to $512^3$ spatial cells. The reported results show that interface smoothing substantially reduces material-coefficient ranks and that the TT--FDTD solution reproduces the full-grid transient fields with pointwise absolute errors on the order of $10^{-4}$ in the examined slices. Compared with conventional FDTD, the tensor representation greatly reduces storage at fine discretizations, although tensor contractions and recompression introduce additional per-step computational cost. These results demonstrate the feasibility and memory--time tradeoff of QTT-accelerated three-dimensional FDTD for large structured-grid simulations.
\end{abstract}

%
%
\section{Introduction}

\IEEEPARstart{T}{he} finite-difference time-domain (FDTD) method
\cite{Yee,MonkSuli1994,TafloveHagness,TafloveHistory2007,TeixeiraFDTDReview} remains one of the most widely used techniques in
computational electromagnetics for broadband analysis of antennas,
microwave circuits, photonic devices, scattering, and wave propagation
problems. Owing to its explicit leapfrog time integration on the
staggered Yee grid, the method is simple to implement, exhibits excellent
parallel scalability, and naturally produces transient and wideband
solutions from a single simulation. Consequently, FDTD has become one of
the standard numerical tools alongside the Method of Moments (MoM) and
the Finite Element Method (FEM) for solving Maxwell's equations.

Despite these advantages, conventional FDTD suffers from the well-known
curse of dimensionality. A three-dimensional computational domain
containing
$N=N_xN_yN_z$
Yee cells requires storage of all electric and magnetic field samples,
leading to
$O(N)$
memory consumption, while every time step performs
$O(N)$
finite-difference operations. For electrically large structures,
multiscale geometries, or high-frequency problems requiring very fine
meshes, the computational cost rapidly becomes prohibitive. Although
modern GPU accelerators and massively parallel computer clusters can
substantially reduce wall-clock time, they do not alter the underlying
linear complexity with respect to the number of spatial unknowns. As
problem sizes continue to grow, memory bandwidth and storage rather than
floating-point throughput increasingly become the dominant computational
bottlenecks.

Low-rank tensor decompositions have recently emerged as an attractive
approach for alleviating this complexity by exploiting the fact that many
high-dimensional physical fields possess substantial algebraic
redundancy. Among them, the Tensor Train (TT) decomposition introduced by
Oseledets and Tyrtyshnikov represents multidimensional arrays as products of small tensor
cores connected through low-rank auxiliary dimensions
\cite{KoldaBader2009,HackbuschKuhn2009,Grasedyck2010,OseledetsTyrtyshnikov2009,OseledetsTT2011,TTCross2010}. When the TT ranks remain moderate,
both memory requirements and algebraic operations can be reduced by
orders of magnitude. Numerous tensor-based techniques have subsequently
been developed for solving integral equations, partial differential
equations, optimization problems, and high-dimensional scientific
computing.

An even more powerful representation for structured Cartesian grids is
the Quantized Tensor Train (QTT) decomposition, in which each spatial
index is expressed through its binary digits before applying the TT
factorization. Rather than representing an
$N_x\times N_y\times N_z$
array directly, QTT reshapes it into a higher-order tensor whose modes
all have dimension two. This binary tensorization transforms dependence
on the physical grid size $N$ into dependence on
$\log_2 N$,
allowing many structured operators—including finite-difference,
shift, identity, and diagonal multiplication operators—to admit low-rank tensor train representations. Consequently, many algebraic
operations scale proportionally to
$O(r^2\log N)$,
where
$r$
denotes the maximum TT rank, instead of
$O(N)$.
QTT has already demonstrated substantial acceleration in numerical
linear algebra and, more recently, in volume integral equation solvers
for computational electromagnetics
\cite{KhoromskijQTT2011,DolgovParabolic2012,KazeevQTT2013,CoronaTTIntegral2017,ChenTT2019,NguyenTTMoM2026}.

Recently, Zhou and Teixeira proposed a TT formulation of the FDTD method
\cite{ZhouTeixeiraTTFDTD}, while Manzini \emph{et al.}
developed related tensor-train finite-difference formulations for
Maxwell's equations \cite{ManziniTTMFDMaxwell}. In these approaches, each
spatial coordinate forms one TT mode, resulting in a three-core
representation of three-dimensional fields. Differential operators are
implemented through one-dimensional finite-difference matrices acting on
the corresponding TT cores, followed by tensor recompression to control
rank growth. These developments demonstrate that tensor decompositions
can substantially reduce memory requirements for FDTD while maintaining
solution accuracy.

The present work follows a different philosophy by employing
\emph{quantized} tensor trains instead of the conventional axis-wise TT
representation. Binary tensorization allows both the electromagnetic
fields and the discrete finite-difference operators on the Yee grid to be
represented directly in QTT matrix-product-operator form. Consequently,
all spatial derivative operations inherit logarithmic complexity with
respect to the total number of grid cells whenever the QTT ranks remain
bounded. 

Since discontinuous dielectric interfaces tend to increase
tensor ranks by introducing high-spatial-frequency components, we further
improve practical compressibility by applying Gaussian smoothing to the
material profile prior to tensorization. This smoothing suppresses
artificial spatial high-frequency content generated by abrupt material transitions
while preserving the physical geometry sufficiently
accurately for FDTD simulations.

The proposed formulation is developed for the scattered-field FDTD
equations on the Yee grid. The scattered electric and magnetic fields,
material-dependent update coefficients, equivalent polarization currents,
and finite-difference curl operators are all represented in compatible
QTT forms. Spatial operations therefore become tensor contractions
between low-rank tensor cores rather than explicit manipulations of full grid
field arrays. The formulation naturally accommodates inhomogeneous
dielectric materials and absorbing boundary conditions while preserving
the explicit leapfrog structure of the conventional FDTD algorithm.

The principal contributions of this work are: 1) a full-vector
three-dimensional scattered-field QTT--FDTD formulation for all six Yee-field
components; 2) QTT matrix-product-operator representations of the staggered
spatial differences, material coefficients, equivalent-current sources, and
boundary treatments; 3) Gaussian regularization of multimaterial interfaces
to limit coefficient-rank growth; and 4) numerical validation on a
heterogeneous anatomical head model and a dielectric sphere, including
accuracy, rank, storage, and per-step timing studies as well as memory use. The remainder of the
paper develops the scattered-field equations and Yee updates, introduces the
QTT representation and operators, describes interface smoothing and the
implementation, and then presents the numerical results and conclusions.


\section{Scattered--Field Maxwell Equations and Equivalent Currents}

Maxwell equations for total fields in the inhomogeneous medium are
\begin{align}
  \nabla \times \gv{H}
  &= \varepsilon(\gv{r})\,\partial_t \gv{E}+\sigma_e(\gv{r})\,\gv{E}, \label{eq:total1}\\
  \nabla \times \gv{E}
  &= -\mu(\gv{r})\,\partial_t \gv{H}
     -\sigma_m(\gv{r})\,\gv{H}. \label{eq:total2}
\end{align}

Substituting $\gv{E} = \gv{E}^i+\gv{E}^s$,
$\gv{H} = \gv{H}^i+\gv{H}^s$ and subtracting the background equations
for $\gv{E}^i,\gv{H}^i$ yields the scattered--field system
\begin{align}
  \nabla\times\gv{H}^s 
    &= \varepsilon(\gv{r})\,\partial_t \gv{E}^s
  + \sigma_e(\gv{r})\,\gv{E}^s+ \gv{J}^{eq}, \label{eq:sfE}\\
  \nabla\times\gv{E}^s &= -\mu(\gv{r})\,\partial_t \gv{H}^s
  - \sigma_m(\gv{r})\,\gv{H}^s
  - \gv{M}^{eq}. \label{eq:sfH}
\end{align}

The terms
\begin{align}
  {
  \gv{J}^{eq}(\gv{r},t)
  = (\varepsilon(\gv{r})-\varepsilon_0)\,\partial_t \gv{E}^i
    + \sigma_e(\gv{r})\,\gv{E}^i
  }
  \label{eq:Jeq}\\[4pt]
  {
  \gv{M}^{eq}(\gv{r},t)
  = (\mu(\gv{r})-\mu_0)\,\partial_t \gv{H}^i
    + \sigma_m(\gv{r})\,\gv{H}^i
  }
  \label{eq:Meq}
\end{align}
are equivalent volume electric and magnetic polarization currents that act as sources for the scattered field. For a non-magnetic dielectric with $\mu(\gv{r})=\mu_0$ and
$\sigma_m(\gv{r})=0$, the magnetic equivalent current vanishes,
$\gv{M}^{eq}=0$, and only $\gv{J}^{eq}$ is present.

\section{Scattered-Field Maxwell Equations in Three Dimensions}

For the general three-dimensional formulation, all field components are
present,
\begin{equation}
\gv{E}
=
\hat{\gv{x}}E_x(x,y,z,t)
+
\hat{\gv{y}}E_y(x,y,z,t)
+
\hat{\gv{z}}E_z(x,y,z,t),
\end{equation}
\begin{equation}
\gv{H}
=
\hat{\gv{x}}H_x(x,y,z,t)
+
\hat{\gv{y}}H_y(x,y,z,t)
+
\hat{\gv{z}}H_z(x,y,z,t),
\end{equation}
and the scattered-field Maxwell equations
\eqref{eq:sfE}--\eqref{eq:sfH}
become
\begin{align}
\varepsilon(\gv r)\,
\partial_t E_x^s
+
\sigma_e(\gv r)E_x^s
&=
\partial_y H_z^s
-
\partial_z H_y^s
-
J_x^{eq},
\label{eq:sf_Ex}
\\
\varepsilon(\gv r)\,
\partial_t E_y^s
+
\sigma_e(\gv r)E_y^s
&=
\partial_z H_x^s
-
\partial_x H_z^s
-
J_y^{eq},
\label{eq:sf_Ey}
\\
\varepsilon(\gv r)\,
\partial_t E_z^s
+
\sigma_e(\gv r)E_z^s
&=
\partial_x H_y^s
-
\partial_y H_x^s
-
J_z^{eq},
\label{eq:sf_Ez}
\\
\mu(\gv r)\,
\partial_t H_x^s
+
\sigma_m(\gv r)H_x^s
&=
\partial_z E_y^s
-
\partial_y E_z^s
-
M_x^{eq},
\label{eq:sf_Hx}
\\
\mu(\gv r)\,
\partial_t H_y^s
+
\sigma_m(\gv r)H_y^s
&=
\partial_x E_z^s
-
\partial_z E_x^s
-
M_y^{eq},
\label{eq:sf_Hy}
\\
\mu(\gv r)\,
\partial_t H_z^s
+
\sigma_m(\gv r)H_z^s
&=
\partial_y E_x^s
-
\partial_x E_y^s
-
M_z^{eq}.
\label{eq:sf_Hz}
\end{align}

\section{Update Equations on the Yee Grid for 3-D FDTD}

Upon use of the standard three-dimensional Yee grid with spacings
$\Delta x$, $\Delta y$, and $\Delta z$, and time discretization
$
t^n=n\Delta t,
$
the electric-field samples
$E_x(i+\tfrac12,j,k)^n$,
$E_y(i,j+\tfrac12,k)^n$,
$E_z(i,j,k+\tfrac12)^n$, 
the magnetic-field samples
$H_x(i,j+\tfrac12,k+\tfrac12)^{n+\frac12}$,
$H_y(i+\tfrac12,j,k+\tfrac12)^{n+\frac12}$, 
$H_z(i+\tfrac12,j+\tfrac12,k)^{n+\frac12}$,
and the equivalent-current samples
$J_x^{eq}$,
$J_y^{eq}$,
$J_z^{eq}$,
$M_x^{eq}$,
$M_y^{eq}$,
and
$M_z^{eq}$
are located at their corresponding Yee positions. Enforcing
\eqref{eq:sf_Ex}--\eqref{eq:sf_Hz}
at the appropriate staggered space--time locations gives
\begin{align}
E_x^{s,n+1}
&=
C_e^aE_x^{s,n}
+
C_e^b
[
\partial_yH_z^{s,n+\frac12}
-
\partial_zH_y^{s,n+\frac12}
-
J_x^{eq,n+\frac12}
],
\label{eq:fdtd_Ex}
\\
E_y^{s,n+1}
&=
C_e^aE_y^{s,n}
+
C_e^b
[
\partial_zH_x^{s,n+\frac12}
-
\partial_xH_z^{s,n+\frac12}
-
J_y^{eq,n+\frac12}
],
\label{eq:fdtd_Ey}
\\
E_z^{s,n+1}
&=
C_e^aE_z^{s,n}
+
C_e^b
[
\partial_xH_y^{s,n+\frac12}
-
\partial_yH_x^{s,n+\frac12}
-
J_z^{eq,n+\frac12}
],
\label{eq:fdtd_Ez}
\\
H_x^{s,n+\frac12}
&=
C_h^aH_x^{s,n-\frac12}
+
C_h^b
[
\partial_zE_y^{s,n}
-
\partial_yE_z^{s,n}
-
M_x^{eq,n}
],
\label{eq:fdtd_Hx}
\\
H_y^{s,n+\frac12}
&=
C_h^aH_y^{s,n-\frac12}
+
C_h^b
[
\partial_xE_z^{s,n}
-
\partial_zE_x^{s,n}
-
M_y^{eq,n}
],
\label{eq:fdtd_Hy}
\\
H_z^{s,n+\frac12}
&=
C_h^aH_z^{s,n-\frac12}
+
C_h^b
[
\partial_yE_x^{s,n}
-
\partial_xE_y^{s,n}
-
M_z^{eq,n}
].
\label{eq:fdtd_Hz}
\end{align}
where the coefficients
$C_e^a$,
$C_e^b$,
$C_h^a$,
$C_h^b$
are functions of
$\varepsilon(\gv r)$,
$\mu(\gv r)$,
$\sigma_e(\gv r)$,
and
$\sigma_m(\gv r)$,

\begin{align}
C_e^a
&=
\frac{\varepsilon-\frac12\sigma_e\Delta t}
{\varepsilon+\frac12\sigma_e\Delta t},
&
C_e^b
&=
\frac{\Delta t}
{\varepsilon+\frac12\sigma_e\Delta t},
\\
C_h^a
&=
\frac{\mu-\frac12\sigma_m\Delta t}
{\mu+\frac12\sigma_m\Delta t},
&
C_h^b
&=
\frac{\Delta t}
{\mu+\frac12\sigma_m\Delta t}.
\end{align}


In PEC regions the total electric field is identically zero. Hence, in the
scattered-field formulation,
\begin{equation}
\gv{E}^{\,s,n+1}=-\gv{E}^{\,i,n+1},
\end{equation}
is enforced at all electric-field Yee locations inside the PEC region
instead of \eqref{eq:fdtd_Ex}--\eqref{eq:fdtd_Ez}, while the
magnetic-field update equations
\eqref{eq:fdtd_Hx}--\eqref{eq:fdtd_Hz} remain unchanged.

Inside dielectric regions, updates are performed according to \eqref{eq:fdtd_Ex}--\eqref{eq:fdtd_Hz} with non-zero equivalent current contributions, while at the Yee grid points in free space, contributions from the equivalent currents vanish. 
At the boundaries of the computational domain, the grid is truncated
using the second-order Mur absorbing boundary conditions (ABC)
\cite{Mur} in their mixed $E$--$H$ formulation. For a boundary whose
outward normal is parallel to the $x$ axis, the continuous boundary
conditions are
\begin{align}
\left(
\partial_x E_y^s
-
c_0^{-1}\partial_t E_y^s
+
\frac{c_0\mu_0}{2}\,
\partial_y H_x^s
\right)\Big|_{x=0}
&=0,
\label{eq:mur3dEy}
\\
\left(
\partial_x E_z^s
-
c_0^{-1}\partial_t E_z^s
+
\frac{c_0\mu_0}{2}\,
\partial_z H_x^s
\right)\Big|_{x=0}
&=0,
\label{eq:mur3dEz}
\end{align}
which involve only first-order spatial and temporal derivatives and make
explicit use of the field samples available from previous time steps.
Analogous expressions are employed on the remaining five boundary faces by cyclic permutation of the coordinate directions, with the normal-derivative signs adjusted on the opposite faces.

As an alternative to the second-order Mur absorbing boundary condition,
the computational domain may be terminated using an ESL perfectly
matched layer (ESL-PML) \cite{BerengerPML,RodenGedneyCPML,JiangCuiPML,ZhouTeixeiraTTFDTD}. 
In the ESL-PML configuration, the physical
FDTD region is surrounded by an absorbing layer in which
coordinate-dependent damping profiles are introduced along the
$x$-, $y$-, and $z$-directions. Denoting these conductivity profiles by
$\sigma_x^{\mathrm{PML}}(\gv{r})$,
$\sigma_y^{\mathrm{PML}}(\gv{r})$, and
$\sigma_z^{\mathrm{PML}}(\gv{r})$, each profile vanishes at the
interface with the physical domain and increases gradually toward the
outer boundary. The resulting modified update coefficients attenuate
the outgoing scattered fields
$\gv{E}^{s}$ and $\gv{H}^{s}$ while minimizing reflections at
the physical-domain--PML interface. In the QTT implementation, the
spatially varying ESL-PML coefficients are represented as diagonal QTT
tensors and applied to the corresponding field tensors
$\mathcal{E}_{\alpha}^{\,n}$ and
$\mathcal{H}_{\alpha}^{\,n+\frac12}$,
$\alpha\in\{x,y,z\}$, through elementwise products followed by QTT
rounding. This preserves the tensorized leapfrog structure of the
interior FDTD updates while providing an absorbing-layer alternative to the
local Mur boundary treatment.

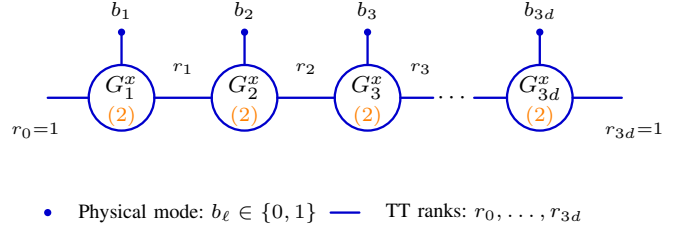
\begin{figure}[!t]
\centering
\resizebox{\columnwidth}{!}{%
\begin{tikzpicture}[
    font=\footnotesize,
    line cap=round,
    line join=round,
    core/.style={
        circle,
        draw=blue!80!black,
        line width=0.8pt,
        minimum size=8mm,
        inner sep=0pt
    },
    bond/.style={
        draw=blue!80!black,
        line width=0.8pt
    },
    phys/.style={
        circle,
        fill=blue!80!black,
        minimum size=2.5pt,
        inner sep=0pt
    }
]


\coordinate (c1) at (0.9,1.2);
\coordinate (c2) at (2.4,1.2);
\coordinate (c3) at (3.9,1.2);
\coordinate (c4) at (6.0,1.2);

\draw[bond] (0,1.2)--(0.5,1.2);
\draw[bond] (1.3,1.2)--(2.0,1.2);
\draw[bond] (2.8,1.2)--(3.5,1.2);
\draw[bond] (4.3,1.2)--(4.7,1.2);
\draw[bond] (5.2,1.2)--(5.6,1.2);
\draw[bond] (6.4,1.2)--(7.0,1.2);

\node[core] (G1) at (c1) {};
\node[core] (G2) at (c2) {};
\node[core] (G3) at (c3) {};
\node[core] (G4) at (c4) {};

\node at ($(c1)+(0,.12)$) {$G_1^x$};
\node[text=orange] at ($(c1)+(0,-.22)$) {\scriptsize$(2)$};

\node at ($(c2)+(0,.12)$) {$G_2^x$};
\node[text=orange] at ($(c2)+(0,-.22)$) {\scriptsize$(2)$};

\node at ($(c3)+(0,.12)$) {$G_3^x$};
\node[text=orange] at ($(c3)+(0,-.22)$) {\scriptsize$(2)$};

\node at ($(c4)+(0,.12)$) {$G_{3d}^x$};
\node[text=orange] at ($(c4)+(0,-.22)$) {\scriptsize$(2)$};

\foreach \x/\lab in {
0.9/$b_1$,
2.4/$b_2$,
3.9/$b_3$,
6.0/$b_{3d}$}
{
    \draw[bond] (\x,1.6)--(\x,2.0);
    \node[phys] at (\x,2.0){};
    \node[above=1pt] at (\x,2.0){\scriptsize\lab};
}

\node at (-0.15,.8){$\scriptstyle r_0=1$};
\node at (1.65,1.55){$\scriptstyle r_1$};
\node at (3.15,1.55){$\scriptstyle r_2$};
\node at (4.55,1.55){$\scriptstyle r_3$};
\node at (7.15,.8){$\scriptstyle r_{3d}=1$};

\node at (4.95,1.2){$\cdots$};

\node[phys] at (0,-0.2){};
\node[anchor=west] at (0.25,-0.2)
{\scriptsize Physical mode: $b_\ell\in\{0,1\}$};

\draw[bond] (3.45,-0.2)--(3.8,-0.2);
\node[anchor=west] at (4.0,-0.2)
{\scriptsize TT ranks: $r_0,\ldots,r_{3d}$};

\end{tikzpicture}%
}
\caption{Tensor-train representation of the electric field $x$-components QTT tensor $\mathcal{E}_x^n$ in \eqref{ex_tensor} with TT order $3d$ in \eqref{ex_tt}.}
\label{fig:qtt_chain}
\end{figure}

\section{QTT Acceleration of the 3D FDTD Scheme}


For simplicity we assume a uniform Yee grid with
$\Delta x=\Delta y=\Delta z$
and
$N_x=N_y=N_z=2^d$
cells in each coordinate direction.
The scattered electric field samples
$
E_x^s\!\left(i+\tfrac12,j,k\right)^n$,
$E_y^s\!\left(i,j+\tfrac12,k\right)^n$,
$E_z^s\!\left(i,j,k+\tfrac12\right)^n$,
are arranged into three-dimensional arrays
$
E^{s,n}_{x,i,j,k}$,
$E^{s,n}_{y,i,j,k}$,
$E^{s,n}_{z,i,j,k}$
where
\(i,j,k=0,\ldots,2^d-1\),
which are then tensorized into
\(3d\)-dimensional QTT tensors
\begin{equation}
\mathcal{E}_x^n
(i_1,\ldots,i_d,
j_1,\ldots,j_d,
k_1,\ldots,k_d),
\label{ex_tensor}
\end{equation}
\begin{equation}
\mathcal{E}_y^n
(i_1,\ldots,i_d,
j_1,\ldots,j_d,
k_1,\ldots,k_d),
\end{equation}
\begin{equation}
\mathcal{E}_z^n
(i_1,\ldots,i_d,
j_1,\ldots,j_d,
k_1,\ldots,k_d)
\in
\mathbb{R}^{2\times\cdots\times2},
\end{equation}
where the binary digits
\(i_\ell,j_\ell,k_\ell\in\{0,1\}\)
encode
\[
i=\sum_{\ell=1}^{d}i_\ell2^{\ell-1},
\qquad
j=\sum_{\ell=1}^{d}j_\ell2^{\ell-1},
\qquad
k=\sum_{\ell=1}^{d}k_\ell2^{\ell-1}.
\]
Denoting the $3d$-dimensional tensor $\mathcal{E}_x^n
(i_1,\ldots,i_d,
j_1,\ldots,j_d,
k_1,\ldots,k_d)$ of the electric field at time step $n$ by $\mathcal{E}_x^n(i,j,k)$, we can state its approximation in the form of the following tensor train
\begin{equation}\label{ex_tt}
\mathcal{E}_x^n(i,j,k)
\approx
G_1^x(b_1)
G_2^x(b_2)
\cdots
G_{3d}^x(b_{3d}),
\end{equation}
where $b_\ell\in\{0,1\},\;
\ell=1,\ldots,3d$ (Fig. \ref{fig:qtt_chain}). For $d=3$, the same TT can be written explicitly as
\begin{equation}
\begin{aligned}
&\mathcal{E}_x^n
(i_1,i_2,i_3,j_1,j_2,j_3,k_1,k_2,k_3)= \\&\mathcal{E}_x^n
(b_1,b_2,b_3,b_4,b_5,b_6,b_7,b_8,b_9) =
\\
&
\sum_{\alpha_1,\ldots,\alpha_8}
G_{1}^x(b_{1},\alpha_{1})
G_{2}^x(\alpha_{1},b_{2},\alpha_{2})
G_{3}^x(\alpha_{2},b_{3},\alpha_{3})
\\
&\quad\times
G_{4}^x(\alpha_{3},b_{4},\alpha_{4})
G_{5}^x(\alpha_{4},b_{5},\alpha_{5})
G_{6}^x(\alpha_{5},b_{6},\alpha_{6})
\\
&\quad\times
G_{7}^x(\alpha_{6},b_{7},\alpha_{7})
G_{8}^x(\alpha_{7},b_{8},\alpha_{8})
G_{9}^x(\alpha_{8},b_{9}).
\end{aligned}
\label{eq:TT_9-dimensional}
\end{equation}
Similarly, the $y$ and $z$ components of the scattered electric field $E_y^s\!\left(i,j+\tfrac12,k\right)^n$,
$E_z^s\!\left(i,j,k+\tfrac12\right)^n$ tensors are treated in their TT form
\begin{equation}
\mathcal{E}_y^n(i,j,k)
\approx
G_1^y(b_1)
G_2^y(b_2)
\cdots
G_{3d}^y(b_{3d}),
\end{equation}
\begin{equation}
\mathcal{E}_z^n(i,j,k)
\approx
G_1^z(b_1)
G_2^z(b_2)
\cdots
G_{3d}^{z}(b_{3d}).
\end{equation}

Scattered magnetic field components $
H_x^s\!\left(i,j+\tfrac12,k+\tfrac12\right)^{n+\frac12}
$, $
H_y^s\!\left(i+\tfrac12,j,k+\tfrac12\right)^{n+\frac12}$
and
$
H_z^s\!\left(i+\tfrac12,j+\tfrac12,k\right)^{n+\frac12}
$
are tensorized in the same manner as the
\(3d\)-dimensional QTT tensors
$
\mathcal{H}_x^{\,n+\frac12}
$,
$\mathcal{H}_y^{\,n+\frac12}$, and
$\mathcal{H}_z^{\,n+\frac12}
$ and approximated in tensor-train form.

\subsection*{QTT Curl Operators}


Let
$
D_x,\;D_y,\;D_z
\in
\mathbb{R}^{2^d\times2^d}
$
be the usual one-dimensional forward-difference matrices (with
appropriate boundary rows) in the $x$-, $y$-, and $z$-directions,
respectively.
Each of these matrices admits an exact low-rank QTT representation with
cores
$\mathsf{D}_x^{(1)},\dots,\mathsf{D}_x^{(d)}$,
$\mathsf{D}_y^{(1)},\dots,\mathsf{D}_y^{(d)}$,
and
$\mathsf{D}_z^{(1)},\dots,\mathsf{D}_z^{(d)}$
acting on the corresponding $d$ binary modes. The induced \emph{three-dimensional} QTT difference operators
\[
\mathbb{D}_x,\;
\mathbb{D}_y,\;
\mathbb{D}_z
:
\mathbb{R}^{2\times\cdots\times2}
\rightarrow
\mathbb{R}^{2\times\cdots\times2}
\]
act on the binary indices associated with the
$x$-, $y$-, and $z$-coordinates, respectively.
Their operator cores can be written compactly as
\begin{equation}
\bigl(\mathbb{D}_x^{(\ell)},\mathbb{D}_y^{(\ell)},\mathbb{D}_z^{(\ell)}\bigr)
=
\begin{cases}
\bigl(\mathsf{D}_x^{(\ell)},\mathsf{I}^{(\ell)},\mathsf{I}^{(\ell)}\bigr),
& \ell=1,\ldots,d,
\\[1ex]
\bigl(\mathsf{I}^{(\ell)},\mathsf{D}_y^{(\ell-d)},\mathsf{I}^{(\ell)}\bigr),
& \ell=d+1,\ldots,2d,
\\[1ex]
\bigl(\mathsf{I}^{(\ell)},\mathsf{I}^{(\ell)},\mathsf{D}_z^{(\ell-2d)}\bigr),
& \ell=2d+1,\ldots,3d,
\end{cases}
\label{eq:QTTDiff3D}
\end{equation}
where $\mathsf{I}^{(\ell)}$ denotes the mode-2 identity-operator core. In a staggered Yee implementation, the symbol $\mathsf D_{\alpha}$ denotes the appropriate forward or backward difference core for the field pair being updated.
Applied to a tensorized field
$\mathcal{X}$,
these operators produce the standard finite differences
\[
(D_x\mathcal{X})_{i,j,k},
\qquad
(D_y\mathcal{X})_{i,j,k},
\qquad
(D_z\mathcal{X})_{i,j,k},
\]
while preserving the logarithmic computational complexity of the QTT
representation. If the difference-operator ranks are bounded, the core contractions for each derivative scale as $O(r^2d)$ for mode size two. The subsequent TT recompression generally introduces an additional rank-dependent cost, commonly of order $O(dr^3)$ for a tensor with comparable ranks \cite{OseledetsTT2011}. Thus, the spatial cost is $O(\operatorname{poly}(r)\log N)$ with $N=2^{3d}$ and is logarithmic in $N$ only while the TT ranks remain bounded.

\subsection*{QTT Form of the Scattered Field 3D Update Equations}

Each coefficient field
$C_{h,\alpha}^a(\gv r)$,
$C_{h,\alpha}^b(\gv r)$,
$C_{e,\alpha}^a(\gv r)$,
and
$C_{e,\alpha}^b(\gv r)$,
where
$\alpha\in\{x,y,z\}$,
is naturally represented as a QTT tensor having the same dimensions as
the corresponding field component.
Denoting these coefficient tensors by
$\mathcal{C}_{h,\alpha}^a$,
$\mathcal{C}_{h,\alpha}^b$,
$\mathcal{C}_{e,\alpha}^a$,
and
$\mathcal{C}_{e,\alpha}^b$,
the equivalent electric and magnetic current tensors by
$\mathcal{J}_{\alpha}^{eq}$ and
$\mathcal{M}_{\alpha}^{eq}$,
and the scattered electric and magnetic field tensors by
$\mathcal{E}_{\alpha}$ and
$\mathcal{H}_{\alpha}$,
the scattered-field FDTD update equations
\eqref{eq:fdtd_Ex}--\eqref{eq:fdtd_Hz}
take the following QTT form.

\begin{align}
\mathcal{E}_x^{\,n+1}
&=
\mathcal{C}_{e,x}^a
\odot
\mathcal{E}_x^{\,n}
\nonumber\\
&\quad+
\mathcal{C}_{e,x}^b
\odot
\left[
\mathbb{D}_y\mathcal{H}_z^{\,n+\frac12}
-
\mathbb{D}_z\mathcal{H}_y^{\,n+\frac12}
-
\mathcal{J}_x^{eq,n+\frac12}
\right],
\label{eq:QTT_Ex}
\\
\mathcal{E}_y^{\,n+1}
&=
\mathcal{C}_{e,y}^a
\odot
\mathcal{E}_y^{\,n}
\nonumber\\
&\quad+
\mathcal{C}_{e,y}^b
\odot
\left[
\mathbb{D}_z\mathcal{H}_x^{\,n+\frac12}
-
\mathbb{D}_x\mathcal{H}_z^{\,n+\frac12}
-
\mathcal{J}_y^{eq,n+\frac12}
\right],
\label{eq:QTT_Ey}
\\
\mathcal{E}_z^{\,n+1}
&=
\mathcal{C}_{e,z}^a
\odot
\mathcal{E}_z^{\,n}
\nonumber\\
&\quad+
\mathcal{C}_{e,z}^b
\odot
\left[
\mathbb{D}_x\mathcal{H}_y^{\,n+\frac12}
-
\mathbb{D}_y\mathcal{H}_x^{\,n+\frac12}
-
\mathcal{J}_z^{eq,n+\frac12}
\right],
\label{eq:QTT_Ez}
\\
\mathcal{H}_x^{\,n+\frac12}
&=
\mathcal{C}_{h,x}^a
\odot
\mathcal{H}_x^{\,n-\frac12}
\nonumber\\
&\quad+
\mathcal{C}_{h,x}^b
\odot
\left[
\mathbb{D}_z\mathcal{E}_y^{\,n}
-
\mathbb{D}_y\mathcal{E}_z^{\,n}
-
\mathcal{M}_x^{eq,n}
\right],
\label{eq:QTT_Hx}
\\
\mathcal{H}_y^{\,n+\frac12}
&=
\mathcal{C}_{h,y}^a
\odot
\mathcal{H}_y^{\,n-\frac12}
\nonumber\\
&\quad+
\mathcal{C}_{h,y}^b
\odot
\left[
\mathbb{D}_x\mathcal{E}_z^{\,n}
-
\mathbb{D}_z\mathcal{E}_x^{\,n}
-
\mathcal{M}_y^{eq,n}
\right],
\label{eq:QTT_Hy}
\\
\mathcal{H}_z^{\,n+\frac12}
&=
\mathcal{C}_{h,z}^a
\odot
\mathcal{H}_z^{\,n-\frac12}
\nonumber\\
&\quad+
\mathcal{C}_{h,z}^b
\odot
\left[
\mathbb{D}_y\mathcal{E}_x^{\,n}
-
\mathbb{D}_x\mathcal{E}_y^{\,n}
-
\mathcal{M}_z^{eq,n}
\right].
\label{eq:QTT_Hz}
\end{align}
Here $\odot$ denotes the elementwise (Hadamard) product between QTT
tensors.
Equations~\eqref{eq:QTT_Ex}--\eqref{eq:QTT_Hz}
constitute the tensorized scattered-field three-dimensional FDTD
formulation. Spatially varying material parameters are represented by
the diagonal QTT coefficient tensors
$\mathcal{C}_{e,\alpha}^{a,b}$
and
$\mathcal{C}_{h,\alpha}^{a,b}$,
while the equivalent source currents are represented by
$\mathcal{J}_{\alpha}^{eq}$
and
$\mathcal{M}_{\alpha}^{eq}$.
All spatial coupling between field components is carried exclusively by
the three QTT difference operators
$\mathbb{D}_x$,
$\mathbb{D}_y$,
and
$\mathbb{D}_z$.

For a uniform grid with $N_x=N_y=N_z=2^d$, the number of QTT cores is $3d=\log_2 N$, where $N=N_xN_yN_z$. With bounded operator and field ranks, the derivative contractions scale linearly with $d$, while additions, Hadamard products, and recompression contribute higher-order polynomial factors in the ranks. The complete update therefore has $O(\operatorname{poly}(r)\log N)$ complexity rather than a rank-independent $O(\log N)$ cost.

To retain a common $N\times N\times N$ QTT layout for all six staggered
Yee components, each component is embedded in a padded array of that size.
Binary support masks $\mathcal M_{E_\alpha}$ and
$\mathcal M_{H_\alpha}$ are one on the physical Yee locations of the
corresponding component and zero on appended ghost planes and other
nonphysical entries. The masks are applied after each update so that padded
samples do not enter subsequent curl operations. Forward and backward
one-dimensional differences are selected consistently with the source and
target Yee locations, even though the compact notation below uses the common
symbols $\mathbb D_x$, $\mathbb D_y$, and $\mathbb D_z$.

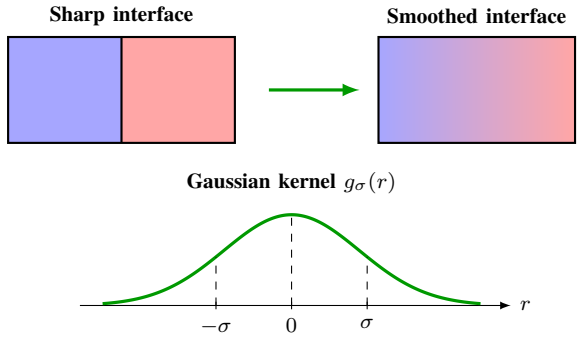
\begin{figure}[!t]
\centering
\begin{tikzpicture}[>=latex,font=\footnotesize]

\node at (1.5,2.68)
{\bfseries Sharp interface};

\node at (6.2,2.68)
{\bfseries Smoothed interface};

\fill[blue!35] (0,1) rectangle (1.5,2.4);
\fill[red!35]  (1.5,1) rectangle (3,2.4);

\draw[thick] (0,1) rectangle (3,2.4);
\draw[thick] (1.5,1)--(1.5,2.4);

\draw[->,very thick,green!60!black]
(3.45,1.7)--(4.65,1.7);

\shade[
    left color=blue!35,
    middle color=white,
    right color=red!35
]
(4.9,1) rectangle (7.5,2.4);

\draw[thick] (4.9,1) rectangle (7.5,2.4);

\node at (3.75,0.48)
{\bfseries Gaussian kernel $g_\sigma(r)$};

\begin{scope}[xshift=3.75cm,yshift=-1.15cm]

\draw[->]
(-2.8,0)--(2.9,0)
node[right] {$r$};

\draw[
    green!60!black,
    very thick,
    samples=150,
    smooth,
    domain=-2.5:2.5
]
plot (\x,{1.20*exp(-\x*\x/1.6)});

\draw[dashed]
(-1,0)--(-1,{1.20*exp(-1/1.6)});

\draw[dashed]
(0,0)--(0,1.20);

\draw[dashed]
(1,0)--(1,{1.20*exp(-1/1.6)});

\draw (-1,0.05)--(-1,-0.05)
node[below] {$-\sigma$};

\draw (0,0.05)--(0,-0.05)
node[below] {$0$};

\draw (1,0.05)--(1,-0.05)
node[below] {$\sigma$};

\end{scope}

\end{tikzpicture}
\caption{Smoothing of a sharp material interface by convolution with a Gaussian kernel.}
\label{fig:gaussian_smoothing}
\end{figure}

\subsection{Smoothing of the material contrasts}
To improve low-rank compressibility of the spatial coefficient tensors and to
reduce high-frequency content introduced by a sharp, staircased material jump,
we replace the binary material indicator by a \emph{Gaussian-smoothed} contrast
computed efficiently in the Fourier domain \cite{NguyenTTMoM2026} (Fig. \ref{fig:gaussian_smoothing}).
Let $m(x,y,z)\in\{0,1\}$ denote the raw (sharp) indicator of the inclusion
($m=1$ inside the object and $m=0$ in the background), sampled on the uniform
grid as $m_{i,j,k}$, $i,j,k=0,\dots,N-1$, with $N=2^d$ and $\Delta=\Delta x=\Delta y = \Delta z$.
We define the smoothed indicator $\tilde m$ via a Gaussian convolution
\begin{align}
  \tilde m(x,y,z) &= (g_\sigma * m)(x,y,z), \nonumber\\
  g_\sigma(x,y,z) &= \frac{1}{(2\pi)^{3/2}\sigma^3}
  \exp\!\left(-\frac{x^2+y^2+z^2}{2\sigma^2}\right).
  \label{eq:gauss_conv_cont}
\end{align}
where $\sigma$ controls the transition thickness. In the discrete periodic setting, the
convolution is evaluated using a 3-D FFT:
\begin{equation}
  \tilde m = \mathcal{F}^{-1}\!\left\{ \widehat{g}_\sigma \odot \widehat{m} \right\},
  \label{eq:gauss_fft_disc}
\end{equation}
where $\widehat{(\cdot)}=\mathcal{F}\{(\cdot)\}$ denotes the 3-D discrete Fourier
transform and $\odot$ is elementwise multiplication. To avoid circular wraparound, the FFT implementation must use sufficient zero padding unless periodic smoothing is intended. For a multimaterial model, the same operation is applied to the indicator of each material class before the permittivity and conductivity fields are reconstructed from the smoothed material weights. The numerical study reports the transition width in grid cells.
\begin{algorithm}[t]
\caption{3-D QTT scattered-field FDTD with Mur's 2nd order ABC}
\label{alg:ttfdtd3d}
\begin{algorithmic}[1]

\Require
QTT difference operators
$\mathbb D_x,\mathbb D_y,\mathbb D_z$;
coefficient tensors
$\mathcal C_{h,\alpha}^{a}$,
$\mathcal C_{h,\alpha}^{b}$,
$\mathcal C_{e,\alpha}^{a}$,
$\mathcal C_{e,\alpha}^{b}$;
equivalent-current tensors
$\mathcal J_{\alpha}^{eq}$,
$\mathcal M_{\alpha}^{eq}$;
Mur's 2nd order face operators;
TT-rounding tolerance $\varepsilon_{\rm TT}$.

\State
Initialize
$\mathcal E_\alpha^{0}$ and
$\mathcal H_\alpha^{-1/2}$,
$\alpha\in\{x,y,z\}$.

\For{$n=0,1,2,\ldots$}

\State
Compute magnetic-field curl terms
\[
\begin{aligned}
\mathcal R_x^H
&=
\mathbb D_y\mathcal E_z^n
-
\mathbb D_z\mathcal E_y^n,
\\
\mathcal R_y^H
&=
\mathbb D_z\mathcal E_x^n
-
\mathbb D_x\mathcal E_z^n,
\\
\mathcal R_z^H
&=
\mathbb D_x\mathcal E_y^n
-
\mathbb D_y\mathcal E_x^n.
\end{aligned}
\]

\State
Update magnetic fields
\[
\mathcal H_x^{n+\frac12}
\leftarrow
\operatorname{round}
\!\left(
\mathcal C_{h,x}^{a}\odot
\mathcal H_x^{n-\frac12}
+
\mathcal C_{h,x}^{b}\odot
\left(
\mathcal R_x^H
-
\mathcal M_x^{eq,n}
\right)
\right)
\]
and similarly for
$\mathcal H_y$ and
$\mathcal H_z$.

\State
Compute electric-field curl terms
\[
\begin{aligned}
\mathcal R_x^E
&=
\mathbb D_y\mathcal H_z^{n+\frac12}
-
\mathbb D_z\mathcal H_y^{n+\frac12},
\\
\mathcal R_y^E
&=
\mathbb D_z\mathcal H_x^{n+\frac12}
-
\mathbb D_x\mathcal H_z^{n+\frac12},
\\
\mathcal R_z^E
&=
\mathbb D_x\mathcal H_y^{n+\frac12}
-
\mathbb D_y\mathcal H_x^{n+\frac12}.
\end{aligned}
\]

\State
Interior electric-field update
\[
\mathcal E_x^{\rm tmp}
\leftarrow
\operatorname{round}
\!\left(
\mathcal C_{e,x}^{a}\odot
\mathcal E_x^{n}
+
\mathcal C_{e,x}^{b}\odot
\left(
\mathcal D_x^E
-
\mathcal J_x^{eq,n+\frac12}
\right)
\right)
\]
and similarly for
$\mathcal E_y$ and
$\mathcal E_z$.

\State
Apply Mur's 2nd order ABC corrections to tangential electric-field components on all six boundary faces.

\State
Set
\begin{equation}
\begin{split}
(
\mathcal E_x^{n+1},
\mathcal E_y^{n+1},
\mathcal E_z^{n+1}
)
\leftarrow
(
\mathcal E_x^{\rm tmp},
\mathcal E_y^{\rm tmp},
\mathcal E_z^{\rm tmp}
) \\
+
\mathcal{B}_{\mathrm{Mur}}
\bigl(\mathcal E^n,\mathcal H^{n+1/2}\bigr).
\end{split}
\end{equation}

\EndFor
\end{algorithmic}
\end{algorithm}
Related work by Zhou and Teixeira and by Manzini \emph{et al.} represents each three-dimensional field component sampled on an $n_1\times n_2\times n_3$ Cartesian grid directly as an order-3 tensor, with one TT core per spatial axis \cite{ZhouTeixeiraTTFDTD,ManziniTTMFDMaxwell}. In that \emph{axis-wise TT} setting, the physical grid sizes $n_1,n_2,n_3$ appear explicitly as the mode sizes, and a one-dimensional difference acts primarily on the core associated with the differentiation direction, followed by TT rounding. By contrast, the QTT approach adopted here reshapes an $N_x\times N_y\times N_z$ field, with $N_x=N_y=N_z=2^d$, into a $3d$-way tensor with binary mode size two. Quantization replaces the large physical modes by many small modes and enables structured matrix-product-operator representations whose dependence on grid size is logarithmic when the associated TT ranks remain moderate.


\section{Numerical Results}
\label{sec:numerical_results}

\subsection{Benchmark Configurations}

The proposed three-dimensional TT--FDTD formulation is evaluated using
two complementary benchmarks: an anatomically realistic multimaterial
human-head model and a homogeneous dielectric sphere. The head model
provides a geometrically and materially complex test case, whereas the
sphere provides a controlled reference for separating the effects of
interface complexity and wave propagation on the tensor ranks.

Both scatterers are embedded in a fixed
$0.4\times0.4\times0.4~\mathrm{m}^{3}$ computational domain and
voxelized on uniform Cartesian grids satisfying
\begin{equation}
N_x=N_y=N_z=N_g=2^d,
\qquad
N_{\mathrm{tot}}=N_g^3=2^{3d},
\label{eq:numerical_grid_size}
\end{equation}
where $N_{\mathrm{tot}}$ denotes the total number of spatial elements.
For the head model, the refinement levels $d=6,7,8,$ and $9$
correspond to grids from $64^3$ to $512^3$, or from
$2.62\times10^5$ to $1.342\times10^8$ spatial elements. The sphere
coefficient-rank study additionally includes $d=5$, corresponding to
a $32^3$ grid.

All field components, material distributions, and update coefficients
are represented using a mode-2 QTT decomposition with the
coordinate-grouped binary ordering
\begin{equation}
[y_1,\ldots,y_d
\,|\,
x_1,\ldots,x_d
\,|\,
z_1,\ldots,z_d].
\label{eq:numerical_qtt_ordering}
\end{equation}
This implementation ordering is a fixed permutation of the
$x$-then-$y$-then-$z$ ordering used in the preceding derivation; the field and
operator cores are permuted consistently, so the physical finite differences
are unchanged. Each three-dimensional QTT tensor therefore contains $3d$ cores,
increasing only from 18 cores at $d=6$ to 27 cores at $d=9$.
The grid spacing decreases from $6.25~\mathrm{mm}$ at $d=6$ to
$0.78125~\mathrm{mm}$ at $d=9$. The time step is selected using a
Courant factor of $0.99$, resulting in approximately 336--2687 time steps over a $4~\mathrm{ns}$ simulation interval for the head study. The sphere transient shown later is advanced to $8~\mathrm{ns}$.

The material-dependent electric-field update tensors
$\mathcal{C}_e^a$ and $\mathcal{C}_e^b$ are compressed using a
relative QTT truncation tolerance of $10^{-4}$. Material-interface
transition widths from zero to four grid cells are examined to
determine the effects of interface smoothing, spatial refinement,
and geometric complexity on the maximum TT/QTT ranks.

\begin{figure}[!t]
    \centering
    \includegraphics[width=\columnwidth]
    {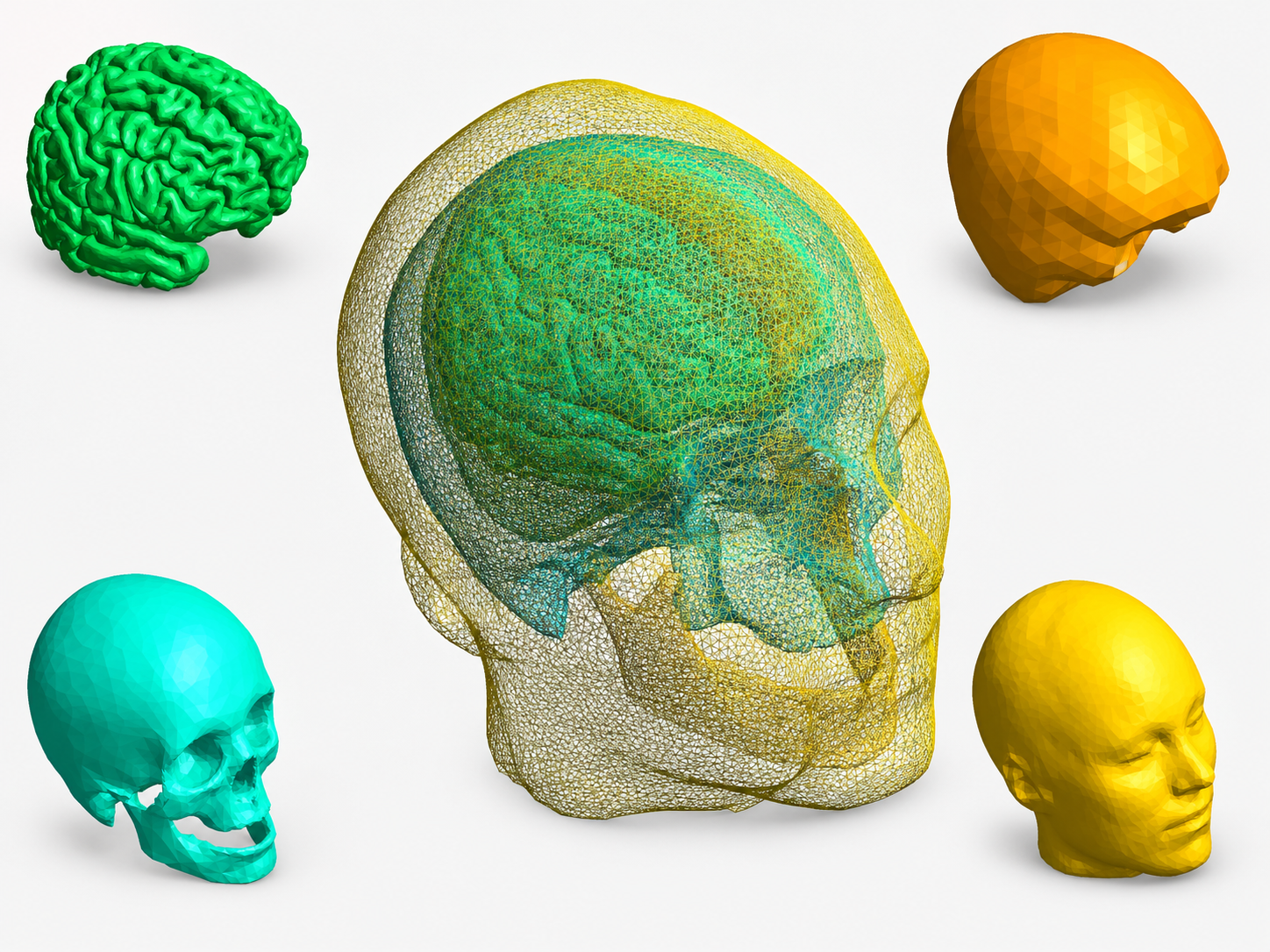}
    \caption{Anatomically realistic multimaterial head model used in
    the three-dimensional TT--FDTD experiments. The geometry contains
    brain, cerebrospinal-fluid, skull, and skin/soft-tissue regions and
    is voxelized on Cartesian grids ranging from $64^3$ to $512^3$
    elements. The original model was provided by Prof. F. Andriulli,
    Politecnico di Torino.}
    \label{fig:head_mesh}
\end{figure}

\begin{table}[!t]
    \centering
    \caption{Electromagnetic properties assigned to the head tissues.}
    \label{tab:head_materials}

    \renewcommand{\arraystretch}{1.30}
    \small

    \begin{tabular*}{\columnwidth}{
        @{\extracolsep{\fill}} l c c @{}
    }
        \hline
        Material
        & $\varepsilon_r$
        & $\sigma~(\mathrm{S/m})$ \\
        \hline
        Air                 & 1  & 0    \\
        Brain               & 50 & 0.60 \\
        Cerebrospinal fluid & 68 & 2.00 \\
        Skull               & 12 & 0.08 \\
        Skin/soft tissue    & 38 & 1.00 \\
        \hline
    \end{tabular*}
\end{table}

The material properties in Table~\ref{tab:head_materials} are representative nondispersive benchmark values informed by standard tissue-property data \cite{Gabriel1996I,Gabriel1996II,Gabriel1996III} and are held fixed at every refinement level. They are used to isolate the numerical effects of refinement and smoothing rather than to model the full dispersive response of tissue over the pulse bandwidth. Consequently, changes in tensor rank
can be attributed to grid refinement and interface regularization
rather than to changes in the physical model.

The homogeneous sphere is centered in the same computational domain
and has radius $0.1343~\mathrm{m}$, relative permittivity
$\varepsilon_r=4$, and conductivity $\sigma=0$. Its dimensions are
comparable to those of the head model, while its material complexity
is confined to a single smooth boundary.

Both models are illuminated by the same linearly polarized Gaussian
plane wave. Let
$\gv{r}=x\hat{\gv{x}}+y\hat{\gv{y}}
+z\hat{\gv{z}}$, and let $\hat{\gv{k}}^{\,i}$ and
$\hat{\gv{p}}^{\,i}$ denote the incident propagation and
polarization directions, respectively, with
$\hat{\gv{k}}^{\,i}\cdot\hat{\gv{p}}^{\,i}=0$. The propagation
direction is
\begin{equation}
\hat{\gv{k}}^{\,i}
=
\sin\theta_i\cos\phi_i\,\hat{\gv{x}}
+
\sin\theta_i\sin\phi_i\,\hat{\gv{y}}
+
\cos\theta_i\,\hat{\gv{z}} .
\label{eq:incident_direction}
\end{equation}
The incident fields are
\begin{align}
\gv{E}^{i}(\gv{r},t)
&=
E_0\hat{\gv{p}}^{\,i}
\exp\!\left[
-\left(
\frac{
t-t_0-\hat{\gv{k}}^{\,i}\cdot\gv{r}/c_0
}{\tau}
\right)^2
\right],
\label{eq:incident_E_3d}
\\
\gv{H}^{i}(\gv{r},t)
&=
\frac{1}{\eta_0}
\hat{\gv{k}}^{\,i}\times\gv{E}^{i}(\gv{r},t),
\label{eq:incident_H_3d}
\end{align}
where $c_0$ and $\eta_0$ are the free-space wave velocity and
impedance, respectively.

For the present experiments,
$\phi_i=45^{\circ}$, $\theta_i=90^{\circ}$, and the electric field is
polarized along the $z$ direction. Hence,
\begin{equation}
\hat{\gv{k}}^{\,i}
=
\frac{\hat{\gv{x}}+\hat{\gv{y}}}{\sqrt{2}},
\qquad
\hat{\gv{p}}^{\,i}=\hat{\gv{z}},
\label{eq:incident_vectors_selected}
\end{equation}
and
\begin{equation}
E_z^{i}(x,y,z,t)
=
E_0
\exp\!\left[
-\left(
\frac{
t-t_0-(x+y)/(\sqrt{2}c_0)
}{\tau}
\right)^2
\right].
\label{eq:incident_Ez_selected}
\end{equation}
The corresponding incident magnetic-field components are
\begin{equation}
H_x^{i}
=
\frac{E_z^{i}}{\sqrt{2}\eta_0},
\qquad
H_y^{i}
=
-\frac{E_z^{i}}{\sqrt{2}\eta_0},
\qquad
H_z^{i}=0.
\label{eq:incident_H_components}
\end{equation}
The source parameters are
\begin{equation}
E_0=1~\mathrm{V/m},
\qquad
t_0=1.0\times10^{-9}~\mathrm{s},
\qquad
\tau=1.5\times10^{-10}~\mathrm{s}.
\label{eq:incident_parameters}
\end{equation}

Although the incident electric field is uniform along $z$, the
scattering problem remains fully three-dimensional because the
material distributions vary in all three coordinates and generate all
scattered-field components. For the nonmagnetic materials considered
here, $\mu=\mu_0$ and $\sigma_m=0$, so the equivalent magnetic current
vanishes. The nonzero electric equivalent current is
\begin{equation}
J_z^{\mathrm{eq},n+\frac{1}{2}}
=
\left(\varepsilon-\varepsilon_0\right)
\frac{E_z^{i,n+1}-E_z^{i,n}}{\Delta t}
+
\sigma_e
\frac{E_z^{i,n+1}+E_z^{i,n}}{2}.
\label{eq:incident_equivalent_current}
\end{equation}

\begin{figure}[!t]
    \centering

    \begin{subfigure}[t]{0.7\columnwidth}
        \centering
        \includegraphics[width=\linewidth]
        {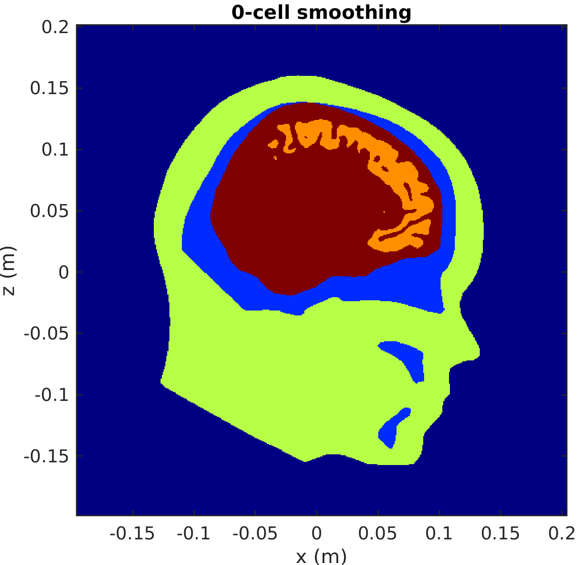}
        \caption{Sharp interfaces.}
        \label{fig:head_epsr_d9_smooth0}
    \end{subfigure}

    \vspace{0.25em}

    \begin{subfigure}[t]{0.70\columnwidth}
        \centering
        \includegraphics[width=\linewidth]
        {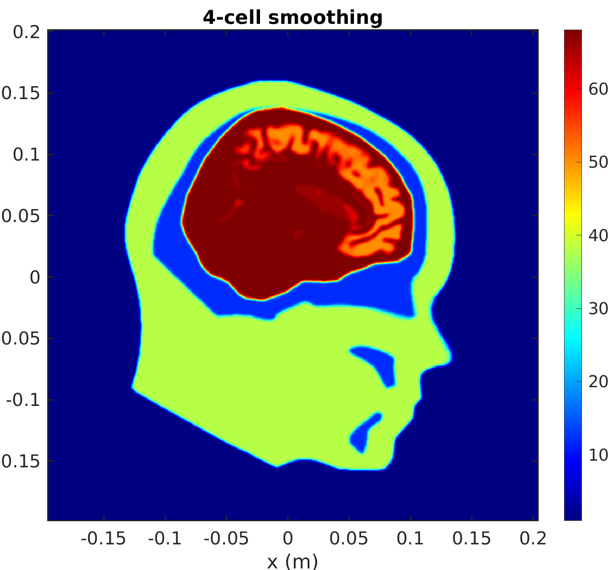}
        \caption{Four-cell transitions.}
        \label{fig:head_epsr_d9_smooth4}
    \end{subfigure}

    \caption{Relative-permittivity maps on the central $y$-slice of
    the $d=9$ head model for (a) sharp material interfaces and
    (b) four-cell interface transitions.}
    \label{fig:head_epsr_d9_smoothing}
\end{figure}

Figure~\ref{fig:head_epsr_d9_smoothing} demonstrates the effect of
interface smoothing on the $d=9$ head model. The sharp material map
contains abrupt voxel-to-voxel transitions at the tissue boundaries,
whereas the four-cell case replaces these discontinuities with smooth
transition layers while retaining the principal anatomical structures
and bulk tissue values. The one-, two-, and three-cell cases are
included in the rank study but omitted from the figure for compactness.
The smoothing suppresses high-spatial-frequency content introduced by
staircasing and improves the QTT compressibility of
$\mathcal{C}_e^a$ and $\mathcal{C}_e^b$.

\subsection{Coefficient and Field-Rank Behavior}

\begin{figure*}[!t]
    \centering

    \begin{subfigure}[t]{0.475\textwidth}
        \centering
        \includegraphics[
            width=\linewidth,
            height=0.75\textwidth
        ]{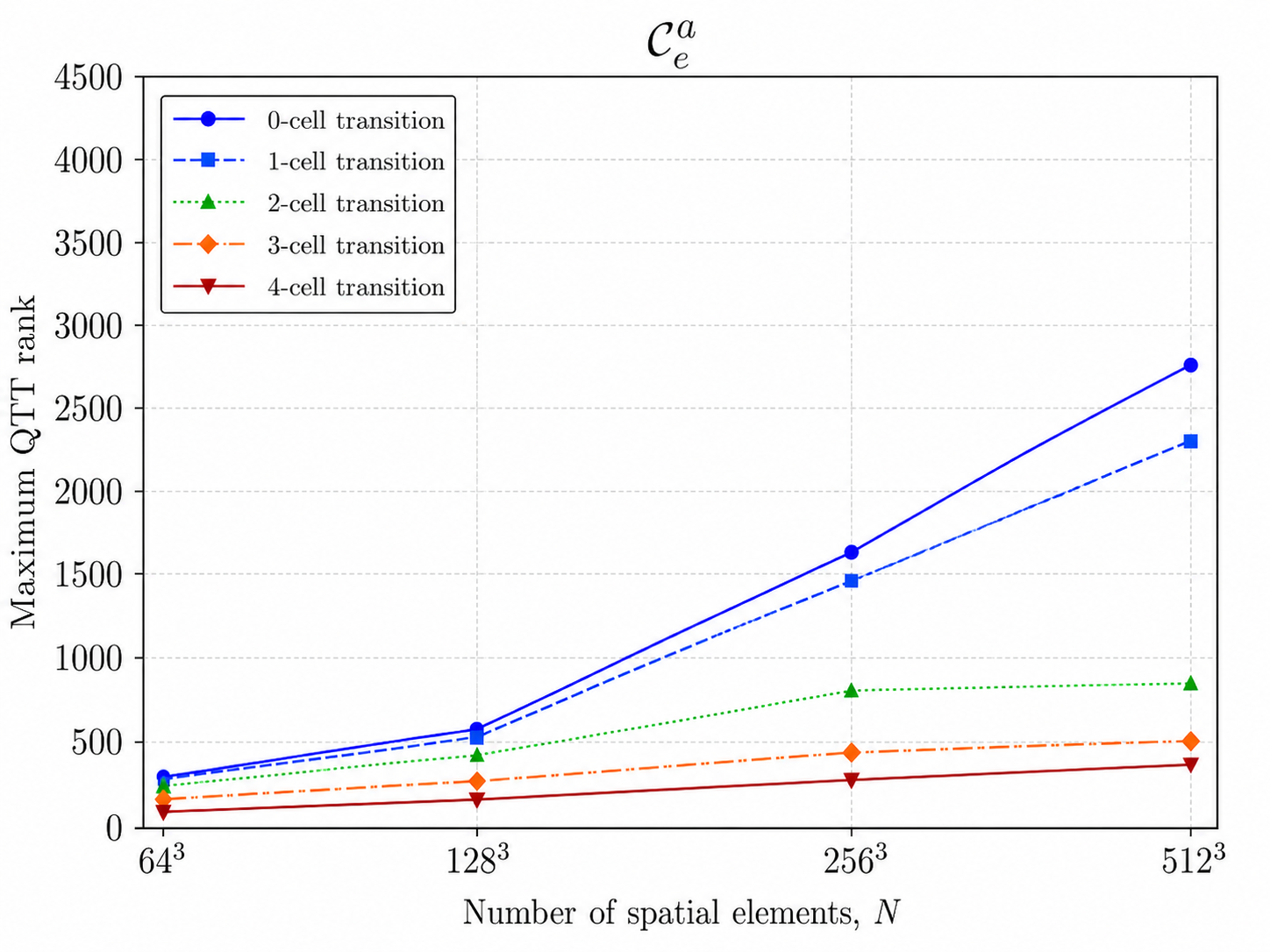}
        \caption{Maximum mode-2 QTT rank of $\mathcal{C}_e^a$
        for the head model.}
        \label{fig:qtt_rank_Ca_material}
    \end{subfigure}
    \hfill
    \begin{subfigure}[t]{0.475\textwidth}
        \centering
        \includegraphics[
            width=\linewidth,
            height=0.75\textwidth
        ]{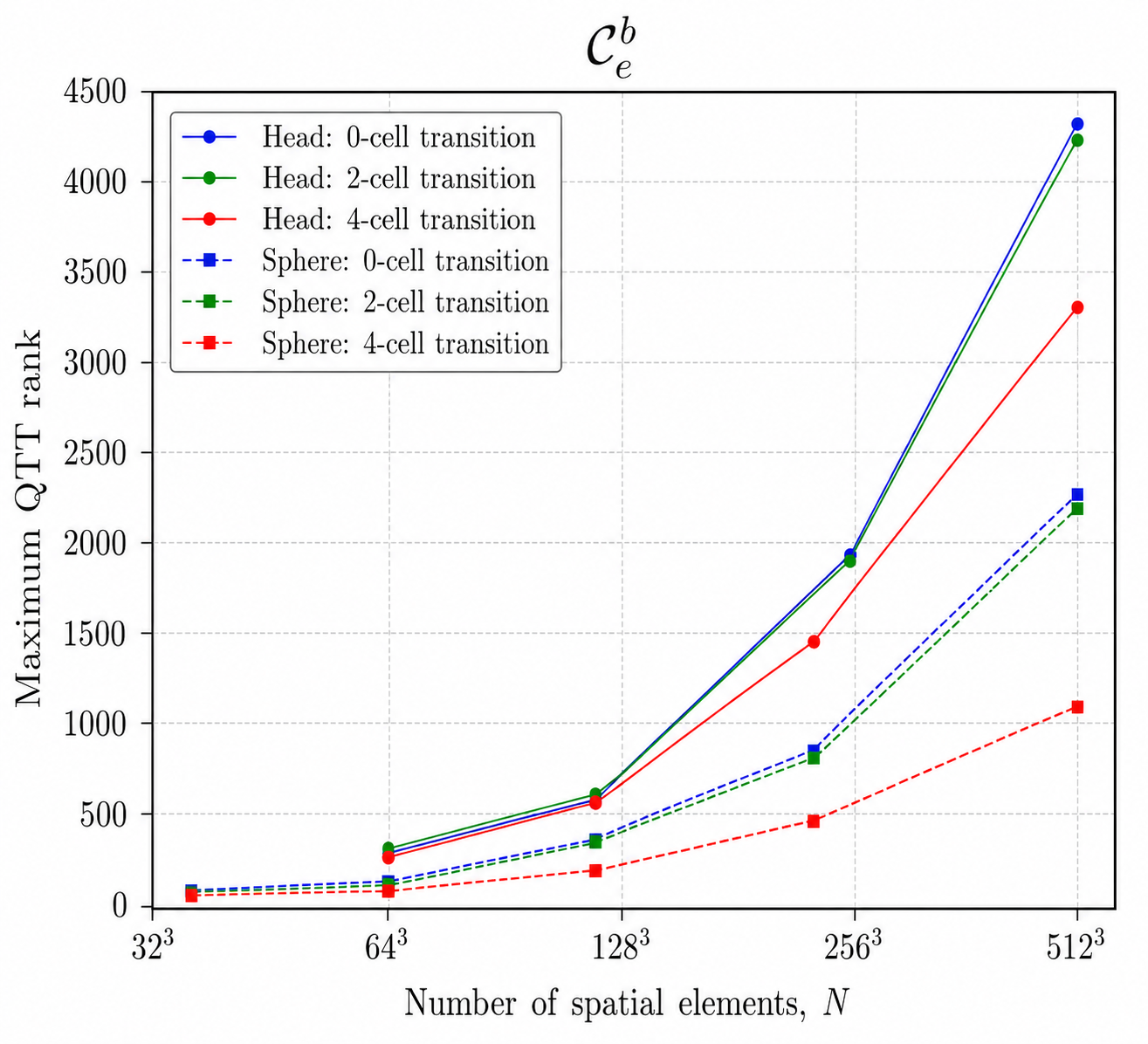}
        \caption{Maximum mode-2 QTT rank of $\mathcal{C}_e^b$
        for the head and sphere models.}
        \label{fig:qtt_rank_Cb_material}
    \end{subfigure}

    \caption{Maximum material-coefficient ranks versus the total
    number of spatial elements. In (a), results are shown for head
    interface-transition widths from zero to four grid cells. In (b),
    solid and dashed curves denote the head and sphere models,
    respectively, for transition widths of 0, 2, and 4 grid cells.
    For the lossless sphere, $\mathcal{C}_e^a=1$ and its QTT rank is
    identically one.}
    \label{fig:material_coefficient_ranks}
\end{figure*}

Figures~\ref{fig:qtt_rank_Ca_material} and
\ref{fig:qtt_rank_Cb_material} show the maximum mode-2 QTT ranks of
the material-dependent electric-field update tensors. For the lossy
head tissues, both $\mathcal{C}_e^a$ and $\mathcal{C}_e^b$ vary
spatially. The sharp-interface rank of $\mathcal{C}_e^a$ increases to
2765 at $d=9$, whereas two- and four-cell smoothing reduce the rank to
829 and 375, respectively. The strong reduction demonstrates the
sensitivity of $\mathcal{C}_e^a$ to voxelized conductivity
transitions.

For the lossless sphere, $\sigma=0$, and therefore
$\mathcal{C}_e^a=1$ throughout the domain. Its QTT rank is consequently
one at every discretization level. The sphere's nontrivial
material-dependent behavior is contained in $\mathcal{C}_e^b$, which
depends on the permittivity distribution.

The head exhibits consistently larger $\mathcal{C}_e^b$ ranks because
its irregular multimaterial interfaces have substantially greater
spatial complexity. At $d=9$, the sharp-interface ranks are 4343 for
the head and 2272 for the sphere. Four-cell smoothing reduces these
values to 3303 and 1083, respectively. This corresponds to a reduction
of approximately 24\% for the head and more than 50\% for the sphere.
The larger reduction for the sphere is expected because its material
complexity is concentrated at one interface, whereas the head retains
multiple irregular internal boundaries after smoothing.

For the time-dependent field-rank study, three consecutive binary
modes are folded into one mode of size eight. This produces a mode-8
TT representation with $d$ cores while preserving the same
$N_{\mathrm{tot}}=8^d=2^{3d}$ spatial samples.

\begin{figure}[!t]
    \centering
    \includegraphics[width=\columnwidth]
    {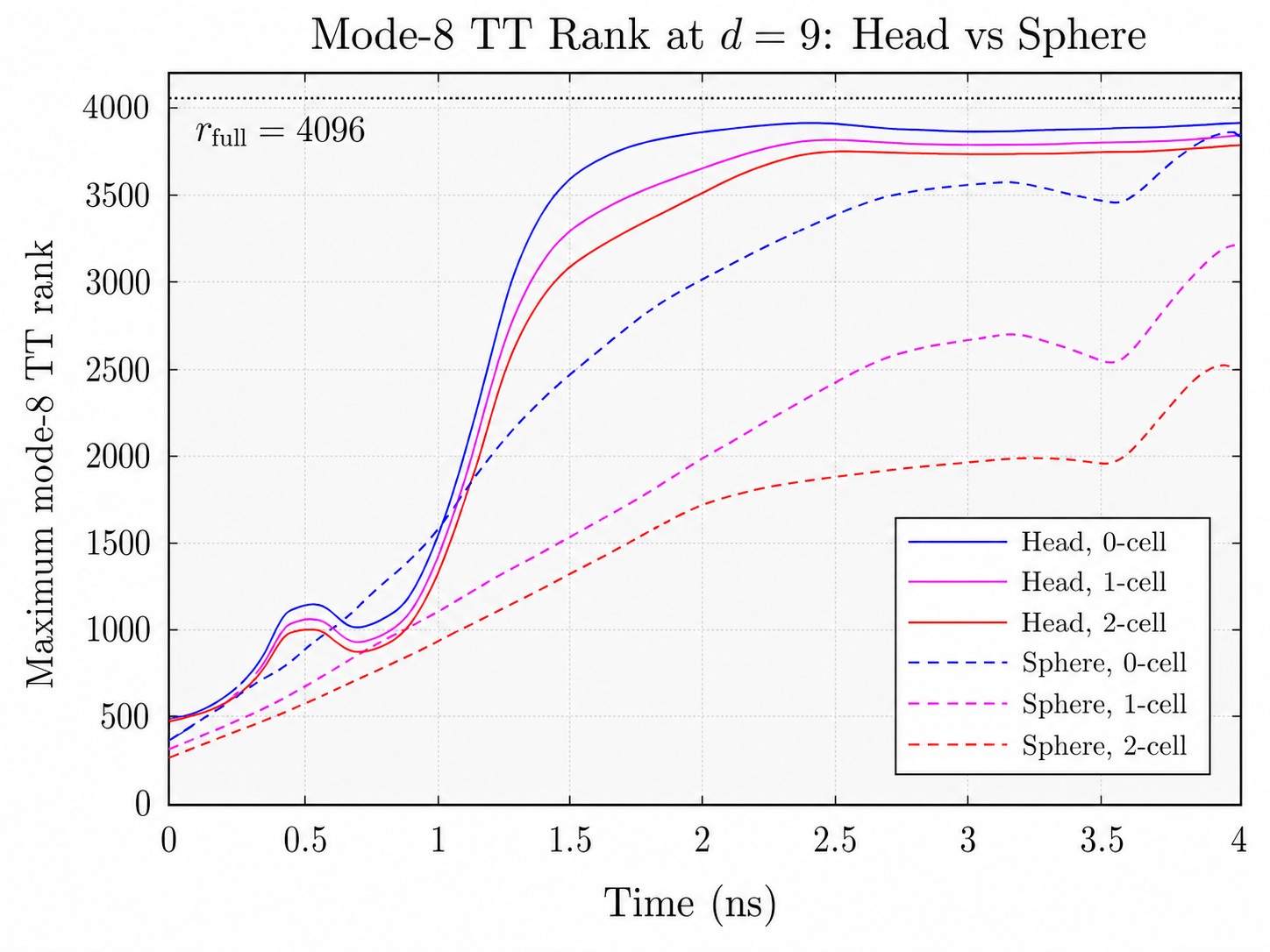}
    \caption{Estimated evolution of the maximum mode-8 TT field rank
    for the head and dielectric-sphere models at $d=9$ during the
    first $4~\mathrm{ns}$. Solid and dashed curves denote the head
    and sphere models, respectively. Results are shown for interface
    transition widths of 0, 1, and 2 grid cells. The horizontal dotted
    line denotes the maximum middle unfolding rank
    $r_{\mathrm{full}}=8^{\lfloor d/2\rfloor}=4096$ for an unconstrained order-$d=9$, mode-8 tensor.}
    \label{fig:head_sphere_mode8_rank}
\end{figure}

Figure~\ref{fig:head_sphere_mode8_rank} compares the estimated
mode-8 TT field-rank evolution of the head and sphere models. The
ranks initially remain moderate while the incident pulse enters the
domain. They subsequently increase as the pulse interacts with the
material regions and the scattered fields occupy a larger portion of
the computational domain.

The head reaches a consistently higher rank regime because its
multilayer anatomical structure, irregular interfaces, and spatially
varying material properties generate a more complex field
distribution. Its curves also remain relatively close as the
transition width increases because smoothing does not remove the
underlying anatomical heterogeneity.

The sphere remains more compressible because its material complexity
is confined to a single boundary. Nevertheless, the unsmoothed sphere
approaches the effective full-grid rank as the propagated field fills
an increasing fraction of the domain. Interface smoothing produces a
more pronounced rank reduction for the sphere, consistent with the
coefficient-rank behavior in
Fig.~\ref{fig:qtt_rank_Cb_material}. Thus, both material complexity
and global wave propagation contribute to the field ranks.


\subsection{Transient-Field Validation for the Head Model}

    \begin{figure*}[!t]
    \centering
    \includegraphics[
        width=0.99\textwidth,
        height=0.85\textheight
    ]{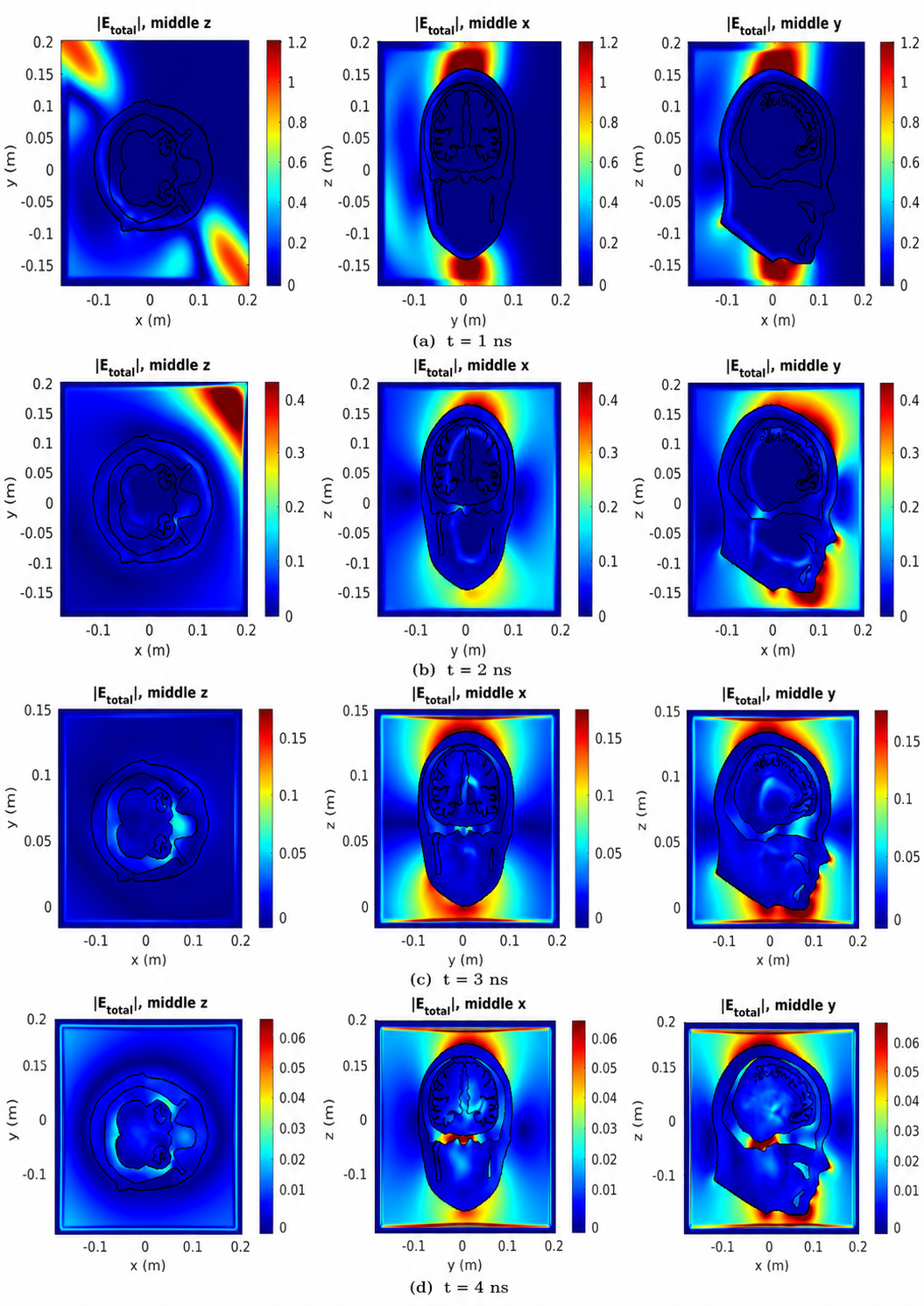}
    \caption{Time evolution of the total electric-field magnitude
    $\left|\gv{E}^{\mathrm{tot}}\right|$ (V/m) in the anatomical head
    model at $d=9$. Each row contains the middle-$z$, middle-$x$,
    and middle-$y$ cross sections at
    (a) $t=1~\mathrm{ns}$,
    (b) $t=2~\mathrm{ns}$,
    (c) $t=3~\mathrm{ns}$, and
    (d) $t=4~\mathrm{ns}$.}
    \label{fig:head_field_time_evolution}
\end{figure*}

Figure~\ref{fig:head_field_time_evolution} shows the evolution of
$\left|\gv{E}^{\mathrm{tot}}\right|$, where
$\gv{E}^{\mathrm{tot}}=\gv{E}^{i}+\gv{E}^{s}$. At
$t=1~\mathrm{ns}$, the incident pulse is concentrated primarily
outside the head and has begun to interact with the outer tissue
boundary. At subsequent times, the field penetrates the heterogeneous
model and develops localized enhancements near tissue interfaces as a
result of transmission, reflection, and internal scattering.

The middle-$x$ and middle-$y$ planes show the progressive
redistribution of the field through the cranial and lower-head regions,
whereas the middle-$z$ plane shows the corresponding transverse
scattering pattern. At later times, the field becomes distributed
throughout the anatomical model and exhibits localized concentrations
associated with its multilayer structure. Because the four rows use
different color-bar limits, the figure is intended to show spatial and
temporal evolution rather than direct comparisons of absolute
magnitude between rows.

The TT--FDTD result is compared with the standard full-grid FDTD
solution using the pointwise absolute field error
\begin{equation}
\varepsilon_E(\gv{r},t)
=
\left|
\gv{E}^{\mathrm{tot}}_{\mathrm{TT}}(\gv{r},t)
-
\gv{E}^{\mathrm{tot}}_{\mathrm{FDTD}}(\gv{r},t)
\right|.
\label{eq:absolute_field_error}
\end{equation}

\begin{figure}[!t]
    \centering
    \makebox[\columnwidth][c]{%
        \includegraphics[width=1.04\columnwidth]
        {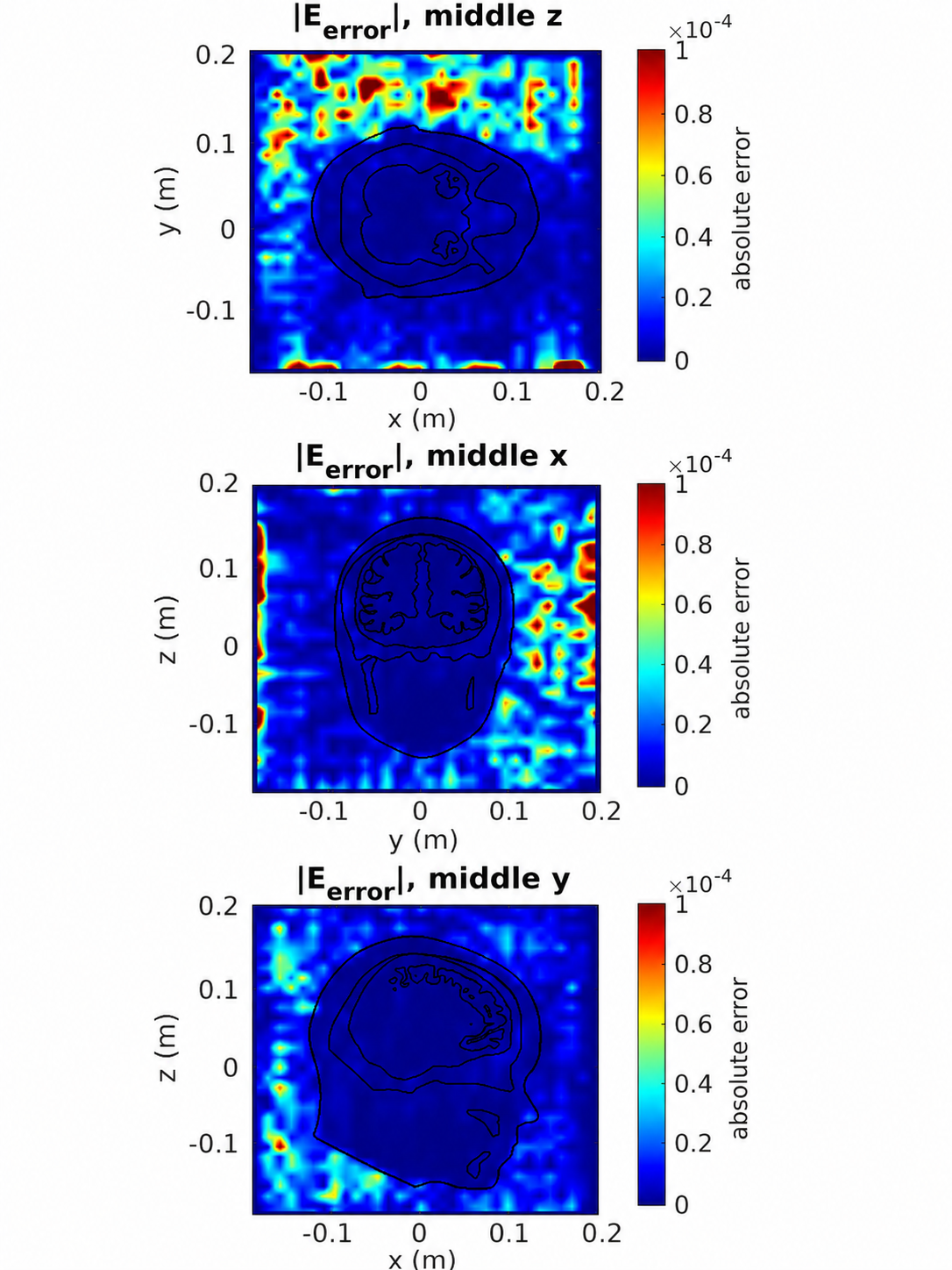}%
    }
    \caption{Pointwise absolute error between the TT--FDTD and
    standard full-grid FDTD solutions at $t=2~\mathrm{ns}$ on the
    middle-$z$, middle-$x$, and middle-$y$ planes of the anatomical
    head model.}
    \label{fig:head_abs_error_2ns}
\end{figure}

Figure~\ref{fig:head_abs_error_2ns} shows that the absolute error
remains on the order of $10^{-4}$ in all three cross sections. The
largest differences occur primarily near the outer computational
region and selected material interfaces, while the interior field
distribution remains in close agreement with the conventional FDTD
reference. This comparison confirms that the TT truncation and
recompression operations preserve the transient solution with high
accuracy.

\subsection{Sphere Response and Computational Scaling}



\begin{figure}[!t]
    \centering
    \includegraphics[
        width=\columnwidth,
        keepaspectratio
    ]{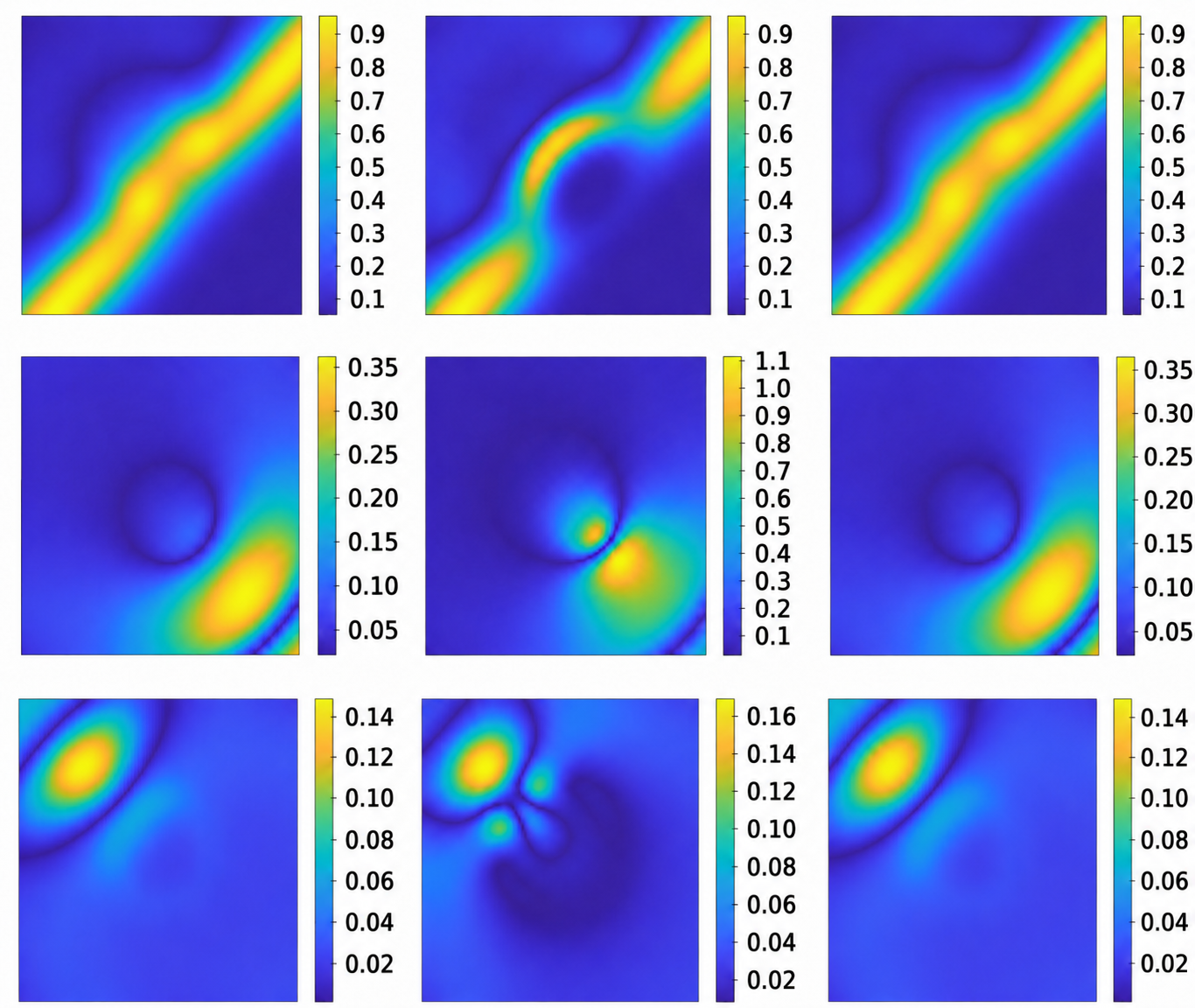}

    \caption{Transient evolution of the total electric-field component
    $\left|E_z^{\mathrm{tot}}(x,y,z_k)\right|$ (V/m) for the dielectric
    sphere at $d=7$, corresponding to $8^7=128^3$ spatial samples.
    The columns correspond to $z_k\in\{32,64,128\}$ from left to
    right, while the rows correspond to $t\approx2$, $4$, and
    $8~\mathrm{ns}$ from top to bottom.}
    \label{fig:sphere_field_slices}
\end{figure}
\begin{figure}[!t]
    \centering
    \includegraphics[width=\columnwidth]
    {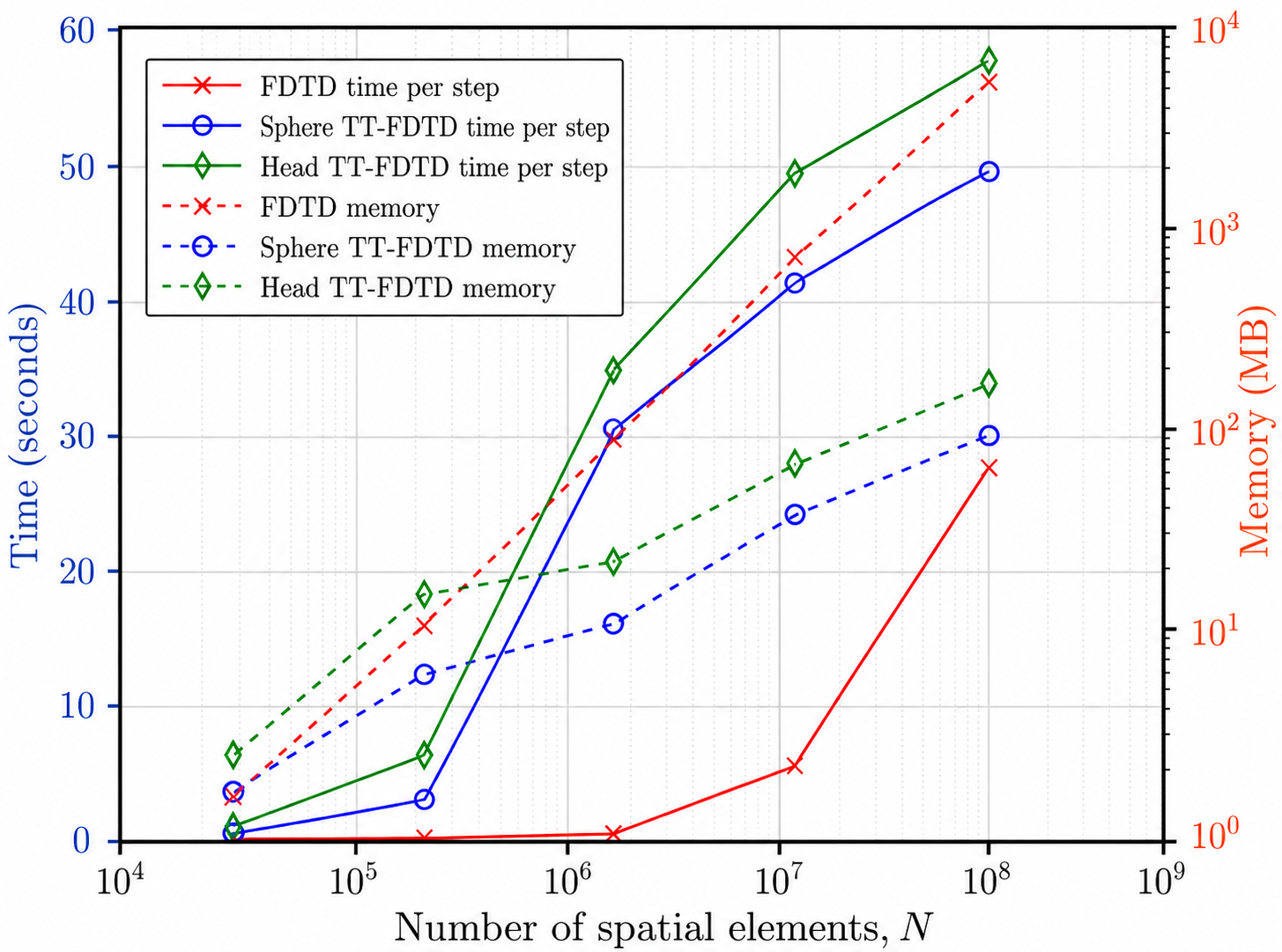}

    \caption{Per-step CPU time and storage memory versus the total
    number of spatial elements $N_{\mathrm{tot}}$ for standard
    full-grid FDTD and TT--FDTD over refinement levels
    $d=5,\ldots,9$. Solid curves denote per-step CPU time and dashed
    curves denote storage memory. The head TT--FDTD curves are
    rank-informed estimates for the anatomical model.}
    \label{fig:time_memory_scaling}
\end{figure}

Figure~\ref{fig:sphere_field_slices} shows representative
$\left|E_z^{\mathrm{tot}}(x,y,z_k)\right|$ slices at three depth
indices and three time instants. At the earliest time, the incident
field remains dominant and the scattered response is beginning to
form. At the intermediate time, the dielectric boundary causes a
clear distortion and redistribution of the field. By
$t\approx8~\mathrm{ns}$, the scattered-field structure is fully
developed and energy has propagated around and away from the sphere.
The controlled geometry makes the effects of the single dielectric
interface easier to identify than in the heterogeneous head model.

Figure~\ref{fig:time_memory_scaling} compares per-step CPU time and
storage memory as the discretization is refined from $d=5$ to $d=9$.
The standard full-grid FDTD memory increases rapidly with
$N_{\mathrm{tot}}$, reaching its largest requirement at the finest
grid. By contrast, both TT--FDTD memory curves grow much more slowly,
demonstrating the storage benefit provided by tensor compression.

The TT--FDTD per-step time is higher than that of standard FDTD because
each update includes tensor contractions and rank truncation. Its
growth, however, remains moderate relative to the reduction in memory.
The head curves lie slightly above the sphere curves because the
multilayer anatomical structure produces larger coefficient and field
ranks. The scaling results therefore demonstrate the principal
memory--time tradeoff of the method: additional tensor-algebra
overhead is exchanged for substantially lower storage requirements at
large three-dimensional discretizations.


\section{Conclusion}
\label{sec:conclusion}

A three-dimensional TT--FDTD formulation was developed for
electromagnetic simulations on large Cartesian grids. The method
represents the electric and magnetic fields, material distributions,
and update coefficients in TT/QTT form, allowing the number of tensor
cores to grow only logarithmically with the number of spatial unknowns.
The formulation was evaluated using both an anatomically heterogeneous
human-head model and a homogeneous dielectric sphere, with
discretizations extending to $512^3$ spatial elements.

The numerical results demonstrate that the material-interface
representation has a strong influence on the coefficient and field
ranks. Interface smoothing substantially reduces the ranks of the
material-dependent update tensors, particularly for the geometrically
simple sphere, whose material complexity is concentrated at a single
boundary. The head model retains higher ranks because of its irregular
multilayer geometry and internal tissue interfaces. Nevertheless, the
TT/QTT representation provides substantial storage savings compared
with conventional full-grid FDTD at the finest discretizations. The
time-dependent results further show that the field ranks are governed
by both material complexity and global wave propagation. Although the
tensor operations introduce additional per-step computational cost,
the method provides a favorable memory--time tradeoff for
large-scale three-dimensional simulations that would otherwise require
prohibitive full-grid storage.



\FloatBarrier

\begin{IEEEbiography}[{\includegraphics[width=1in,height=1.25in,clip,keepaspectratio]{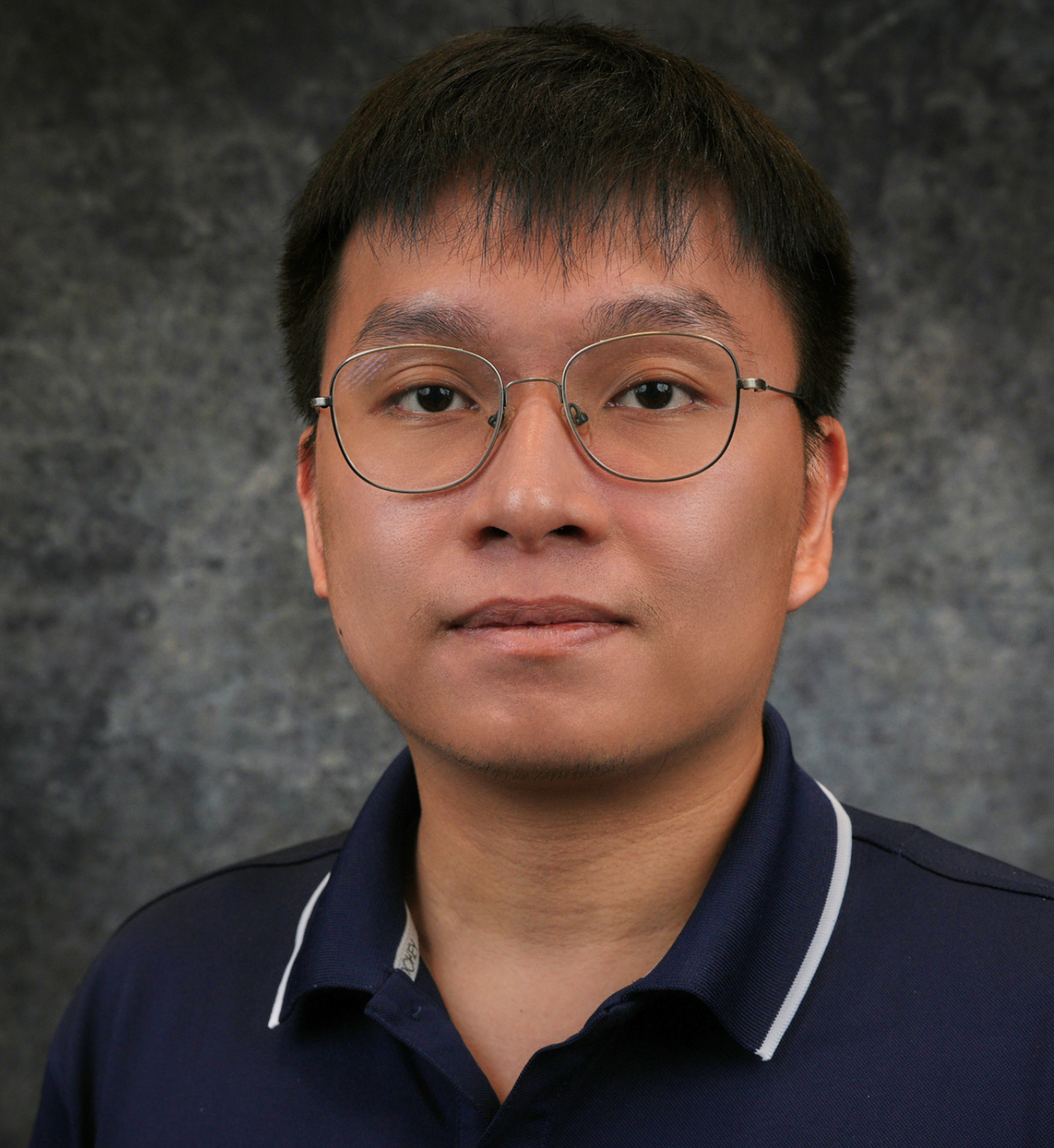}}]{Chris Nguyen}
(M'25)  received the B.Sc. degree from the University of Manitoba, Canada, in 2024.
He is currently an M.Sc. student in the same institution. His research interests are in computational electromagnetics, tensor train decomposition methods, as well as quantum and high-performance computing. He was recipient of the Best Student Paper Award at the 2025 IEEE Conference on Electrical Performance of Electronic Packaging and Systems (EPEPS) for his work on tensor train decompositions in computational electromagnetics.
\end{IEEEbiography}
\begin{IEEEbiography}
[{\includegraphics[width=1in,height=1.25in,clip,keepaspectratio]{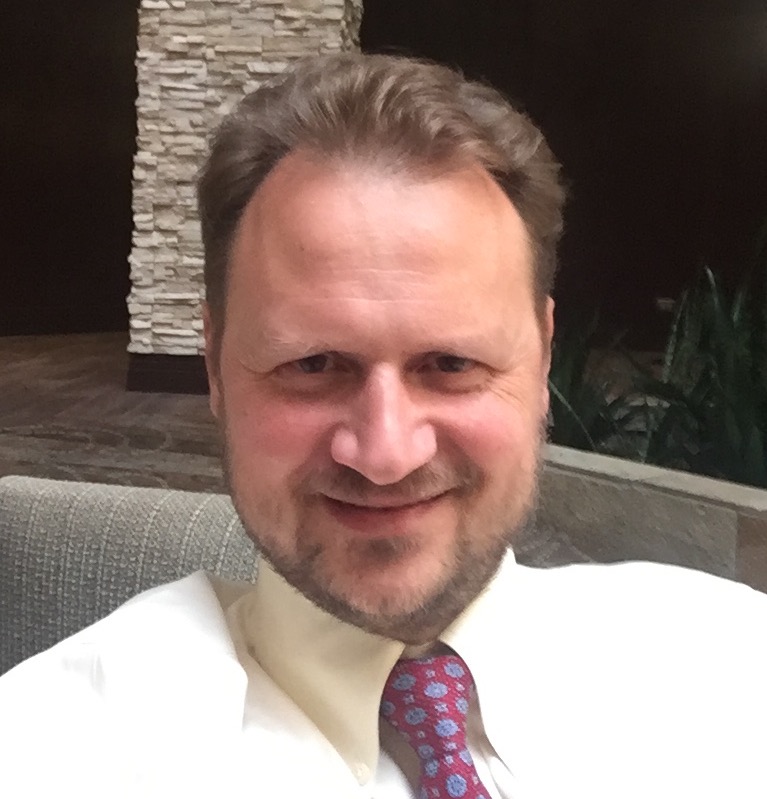}}]
{Vladimir I. Okhmatovski} (M'99--SM'09) was born in Moscow, Russia, in 1974. He received the M.S. degree (with distinction) in radiophysics and the Ph.D. degree in antennas and microwave circuits from the Moscow Power Engineering Institute, Moscow, Russia, in 1996 and 1997, respectively.

From 1997 to 1998, he was an Assistant Professor with the Moscow Power Engineering Institute. He subsequently held postdoctoral appointments with the National Technical University of Athens, Greece, and the University of Illinois at Urbana--Champaign, USA. After serving as a Senior Member of Technical Staff with Cadence Design Systems and as an independent consultant, he joined the Department of Electrical and Computer Engineering, University of Manitoba, Winnipeg, MB, Canada, in 2004, where he is currently a Full Professor. His research interests include fast computational electromagnetics, high-performance computing, interconnect modeling, and inverse problems. He has co-authored a book, a book chapter, and more than 170 technical papers and patents in computational electromagnetics.

Prof. Okhmatovski has served on the Technical Program Review Committee of the IEEE MTT-S International Microwave Symposium since 2017 and on the IEEE MTT-S Technical Committee on Field Theory and Computational Electromagnetics since 2020. He was the Technical Program Co-Chair of the 2021 Applied and Computational Electromagnetics Society Symposium and the General Co-Chair of the 2025 ACES Symposium. He has also served as Chair and Vice-Chair of the IEEE Antennas and Propagation Society Membership and Benefits Committee since 2018.

His awards include the 2017 Intel Corporate Research Council Outstanding Researcher Award, the Outstanding ACES Journal Paper Award (2007), the Best Paper Award at the 3rd Electronic Packaging Technology Conference (2001), the Best Young Scientist Report Award at the VI International Conference on Mathematical Methods in Electromagnetic Theory (1996), and scholarships from the Government of the Russian Federation (1995) and the President of the Russian Federation (1996).

Prof. Okhmatovski is a Registered Professional Engineer in the Province of Manitoba, Canada, and a Fellow of the Applied Computational Electromagnetics Society.
\end{IEEEbiography}

\end{document}